\documentclass[structabstract]{aa}

\usepackage{url}
\usepackage[breaklinks=true]{hyperref}
\usepackage{twoopt}
\usepackage[english]{babel}          

\usepackage{natbib}
\bibpunct{(}{)}{;}{a}{}{,} 

\usepackage[utf8]{inputenc}
\usepackage[normalem]{ulem}
\usepackage{siunitx}
\usepackage{amsmath, amssymb, amsfonts}
\usepackage[farskip=0pt]{subfig}

\usepackage{multirow,bigstrut,ctable}
\usepackage[normalem]{ulem}
\usepackage{rotating}
\usepackage{threeparttable}

\def \deg         {\text{$^{\circ}$}}

\def \arcsec      {\text{$^{\prime\prime}$}}

\def \mjybeam     {mJy\,beam$^{-1}$}
\def \mujybeam    {$\mathrm{\mu}$Jy\,beam$^{-1}$}

\newcommand{\beam}[2]{{#1}\arcsec$\times${#2}\arcsec}

\begin{document}

\title{Systematic effects in LOFAR data: A unified calibration strategy}
\titlerunning{LOFAR calibration strategy}

\author{F.~de~Gasperin\inst{1,2}
\and T.~J.~Dijkema\inst{3}
\and A.~Drabent\inst{4}
\and M.~Mevius\inst{3}
\and D.~Rafferty\inst{1}
\and R.~van~Weeren\inst{2}
\and M.~Br\"uggen\inst{1}
\and J.~R.~Callingham\inst{3}
\and K.~L.~Emig\inst{2}
\and G.~Heald\inst{5}
\and H.~T.~Intema\inst{2}
\and L.~K.~Morabito\inst{6}
\and A.~R.~Offringa\inst{3}
\and R.~Oonk\inst{2,3,7}
\and E.~Orr\`u\inst{3}
\and H.~R\"ottgering\inst{2}
\and J.~Sabater\inst{9}
\and T.~Shimwell\inst{2,3}
\and A.~Shulevski\inst{3}
\and W.~Williams\inst{8}
}
\authorrunning{F.~de~Gasperin et al.}

\institute{
Hamburger Sternwarte, Universit\"at Hamburg, Gojenbergsweg 112, 21029, Hamburg, Germany, \email{fdg@hs.uni-hamburg.de}
\and Leiden Observatory, Leiden University, P.O.Box 9513, 2300 RA, Leiden, The Netherlands
\and ASTRON, the Netherlands Institute for Radio Astronomy, Postbus 2, 7990 AA, Dwingeloo, The Netherlands
\and Th\"uringer Landessternwarte, Sternwarte 5, 07778, Tautenburg, Germany
\and CSIRO Astronomy and Space Science, PO Box 1130, Bentley WA 6102, Australia
\and Astrophysics, University of Oxford, Denys Wilkinson Building, Keble Road, Oxford OX1 3RH, UK
\and SURFsara, P.O. Box 94613, 1090 GP Amsterdam, the Netherlands
\and Centre for Astrophysics Research, School of Physics, Astronomy and Mathematics, University of Hertfordshire, College Lane, Hatfield AL10 9AB, UK
\and Institute for Astronomy, University of Edinburgh, Royal Observatory, Blackford Hill, Edinburgh, EH9 3HJ, UK}

\date{Received ... / Accepted ...}

\abstract
{New generation low-frequency telescopes are exploring a new parameter space in terms of depth and resolution. The data taken with these interferometers, for example with the LOw Frequency ARray (LOFAR), are often calibrated in a low signal-to-noise ratio regime and the removal of critical systematic effects is challenging. The process requires an understanding of their origin and properties.}
{In this paper we describe the major systematic effects inherent to next generation low-frequency telescopes, such as LOFAR. With this knowledge, we introduce a data processing pipeline that is able to isolate and correct these systematic effects. The pipeline will be used to calibrate calibrator observations as the first step of a full data reduction process.}
{We processed two LOFAR observations of the calibrator 3C\,196: the first using the Low Band Antenna (LBA) system at 42--66 MHz and the second using the High Band Antenna (HBA) system at 115--189 MHz.}
{We were able to isolate and correct for the effects of clock drift, polarisation misalignment, ionospheric delay, Faraday rotation, ionospheric scintillation, beam shape, and bandpass. The designed calibration strategy produced the deepest image to date at 54 MHz. The image has been used to confirm that the spectral energy distribution of the average radio source population tends to flatten at low frequencies.}
{We prove that LOFAR systematic effects can be described by a relatively small number of parameters. Furthermore, the identification of these parameters is fundamental to reducing the degrees of freedom when the calibration is carried out on fields that are not dominated by a strong calibrator.}

\maketitle

\section{Introduction}
\label{sec:introduction}

Observing at low radio frequencies ($<1$~GHz) has been a long-standing challenge because of the strength of the systematic effects corrupting the data. However, this poorly explored observational window encodes crucial information for a number of scientific cases. Some examples are the study of low-energy/aged cosmic-ray electrons in galaxies, galaxy clusters \citep[e.g.][]{Hoang2017,deGasperin2017}, and active galactic nuclei \citep[e.g.][]{bgmv16, Harwood2016}; the detection of low-frequency radio recombination lines \citep[e.g.][Emig et al., this issue]{Morabito2014a}; the hunt for high-$z$ radio galaxies \citep[e.g.][]{Saxena2018}, or the exploration of the epoch of reionisation \citep[e.g.][]{Patil2017}.

To achieve high dynamic range and resolution, low-frequency data reduction employs complex schemes aimed to track and correct a number of systematic effects \citep{Williams2016, vanWeeren2016b, Tasse2018}. In a high signal-to-noise ratio (S/N) regime, a brute force calibration that has no assumptions concerning the systematic effects that it aims to correct for, is satisfactory. However, at low frequency the sky temperature is high and observations are plagued by systematic corruptions mainly caused by ionospheric disturbances. These corruptions are time, frequency, and direction dependent; therefore, in the low S/N regime, calibration of these effects is challenging. An effective way to tackle this problem is to reduce the number of free parameters in the calibration by incorporating the (i) time, (ii) frequency, (iii) polarisation, and (iv) spatial coherency of the systematic effects for which we aim to solve.

A fundamental step in this process is to identify as many systematic effects as possible by understanding the response of the telescope when observing bright, compact, and well-characterised sources (i.e. calibrators). Once identified, these effects can be physically characterised to determine their frequency dependency, time/space coherency scale, and polarisation properties. The effects can then be isolated and removed to facilitate the characterisation of higher order effects. Furthermore, in cases in which a calibrator is observed before and after the target fields, all effects that are time and direction independent can be corrected on the target field using the high S/N calibrator solutions; this is a conventional approach in radio astronomy. For phased array such as the LOw Frequency ARray (LOFAR), certain observations can be carried out simultaneously pointing one or more target fields and the calibrator. In such cases, the only requirement for a solution to be transferred from the calibrator to the target is to correct for a direction-independent systematic effect.

The LOw Frequency ARray \citep[LOFAR;][]{VanHaarlem2013} is a radio interferometer capable of observing at very low frequencies ($10-240$~MHz). Each LOFAR station is composed of two sets of antennas: the Low Band Antennas (LBA) operating between 10 and 90 MHz, and the High Band Antennas (HBA) operating between 110 and 250 MHz. Currently, LOFAR is composed of 24 core stations (CSs), 14 remote stations  (RSs), and 13 international stations (ISs). The CSs are spread across a region of radius $\sim2$~km and provide a large number of short baselines. The RSs are located within 90~km from the core and provide a resolution of $\sim15\arcsec$ at 54~MHz and of $\sim5\arcsec$ at 150~MHz. The ISs provide more than another factor of 10 in resolution.

\begin{figure}
\centering
 \includegraphics[width=0.5\textwidth]{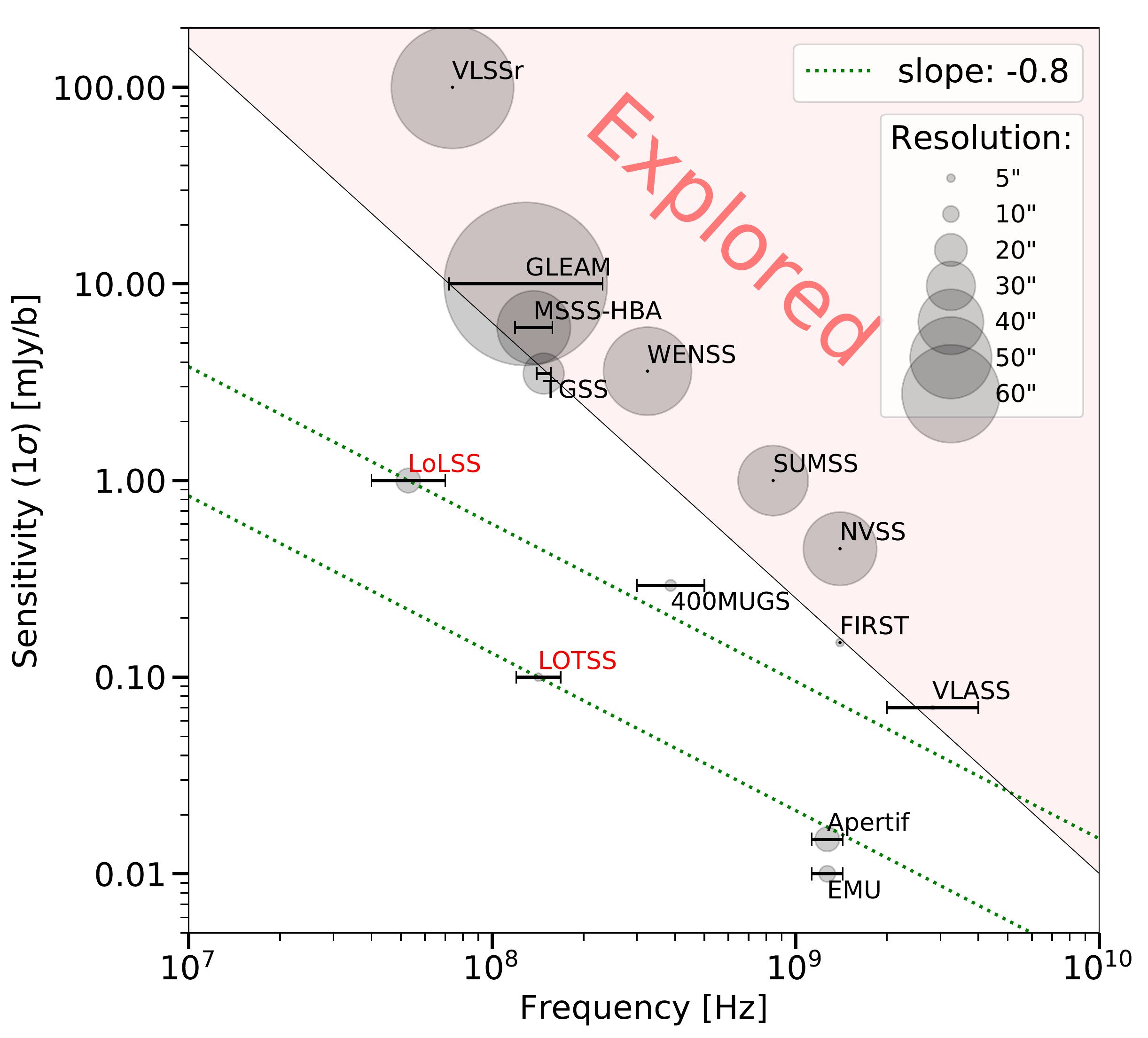}
 \caption{Sensitivity comparison among a number of current, ongoing, and planned large-area radio surveys. The diameters of grey circles are proportional to the survey beam size as shown in the upper right corner. For wide band surveys we show the frequency coverage using horizontal lines.  References: GLEAM \citep[GaLactic and Extragalactic All-sky Murchison Widefield Array survey;][]{Hurley-Walker2017}; MSSS-HBA \citep[LOFAR Multi-frequency Snapshot Sky Survey;][]{Heald2015}; TGSS ADR1 \citep[TIFR GMRT Sky Survey - Alternative Data Release 1;][]{Intema2017}; VLSSr \citep[VLA Low-frequency Sky Survey redux;][]{Lane2014}; FIRST \citep[Faint Images of the Radio Sky at Twenty Centimetres;][]{Becker1995}; NVSS \citep[1.4 GHz NRAO VLA Sky Survey;][]{Condon1998}; WENSS \citep[The Westerbork Northern Sky Survey;][]{Rengelink1997}; SUMSS \citep[Sydney University Molonglo Sky Survey;][]{Bock1999a}; 400MUGS (400 MHz Upgraded GMRT Survey; de Gasperin et al. in prep.); EMU \citep[Evolutionary Map of the Universe;][]{Norris2011}; Apertif \citep[][]{Rottgering2012}; VLASS (VLA Sky Survey; Lacy et al. in prep.); LOTSS \citep[LOFAR Two-metre Sky Survey;][]{Shimwell2016a}; LoLSS (LOFAR LBA Sky Survey; de Gasperin et al. in prep.)}
 \label{fig:surveys}
\end{figure}

One of the primary ambitions of LOFAR is to perform ground-breaking imaging surveys \citep{Rottgering2012}:
\begin{itemize}
 \item \textbf{LoTSS} \citep[LOFAR Two-metre Sky Survey;][]{Shimwell2016a}, is a sensitive, high-resolution survey of the northern sky in the frequency range 120--168 MHz. The survey is currently ongoing and the first full-quality data release of 424 sq. deg. incorporating a direction-dependent error correction has been published \citep{Shimwell2018}. The survey aims to cover the entire northern sky with a depth of 100~\mujybeam and a resolution of 5\arcsec.
 \item \textbf{LoLSS} (LOFAR LBA Sky Survey; de Gasperin et al. in prep.), is the ultra-low-frequency counterpart of LoTSS and will produce an unprecedented view of the sky at 40--70 MHz. The survey has demonstrated capability to achieve 15\arcsec{} resolution with an rms noise of $\sim1$~\mjybeam. The survey records data only from the Dutch stations and it is currently ongoing.
\end{itemize}
Both surveys will be complemented by deeper tiers of observation over smaller sky areas. A comparison between these and other radio surveys is shown in Fig.~\ref{fig:surveys}. As demonstrated in the plot, the aim of the LOFAR survey programme is to push the boundaries of the low- and ultra-low-frequency exploration of the sky, improving by two orders of magnitude in sensitivity and one order of magnitude in angular resolution over previous experiments at comparable wavelengths.

In this work we describe the major sources of systematic effects present in the LOFAR data and outline a calibration scheme that can be used for both LBA and HBA data sets. In Sec.~\ref{sec:systematics} we describe the systematic effects present in LOFAR data. In Sec.~\ref{sec:observations} we present the LBA and HBA observations we will use as test data sets. In Sec.~\ref{sec:strategy} we outline the calibration strategy step by step, while the resulting images are presented in Sec.~\ref{sec:images}.

\section{Systematic effects}
\label{sec:systematics}

In this section, we summarise the most important systematic effects that are present in LOFAR data. In order to describe these effects we use the radio interferometer measurement equation (RIME) formalism, which is described in detail in the first two papers of ``Revisiting the radio interferometer measurement equation'' \citep{Smirnov2011a,Smirnov2011b}. In the RIME formalism every systematic effect corresponds to an operator expressed by a $2\times2$ complex matrix. In line with \cite{Smirnov2011a}, we use the Jones matrix convention \citep{Jones1941} initially adopted by \cite{Hamaker2000} as opposed to the older $4\times4$ Muller matrix convention of the first RIME paper \citep{Hamaker1996}. In this formalism a scalar corresponds to an effect that applies to both polarisations independently. A diagonal matrix describes a polarisation-dependent effect without leakage terms. Effects with non-zero off-diagonal terms (e.g. Faraday rotation) represent a transfer of signal from one polarisation to another.

An important concept that we recall from the RIME formalism is the Jones chain. If multiple effects are present along the signal path of an observation, then this corresponds to a series of matrix multiplication called a Jones chain. The order of terms in a Jones chain is the same as the physical order in which the effects occur along the signal path. It is important to note that matrices can commute only under certain circumstances\footnote{1. Scalars commute with everything. 2. Diagonal matrices commute among themselves. 3. Rotation matrices commute among themselves.}, therefore the order in which  we apply them matters. 

A summary of the properties of the systematic effects considered in this paper is present in Table~\ref{tab:effects}. In the following sections we describe each of these effects in detail.

\begin{table*}
\centering
\begin{threeparttable}
\begin{tabular}{lccccc}
Systematic effect & Type of & Ph/Amp/Both\tnote{b} & Frequency & Direction & Time \\
 & Jones matrix\tnote{a} &  & dependency & dependent? & dependent? \\
\hline
Clock drift & Scalar & Ph & $\propto \nu$ & No & Yes (many seconds) \\
Polarisation alignment & Diagonal & Ph & $\propto \nu$ & No & No \\
Ionosphere - 1st ord. (dispersive delay) & Scalar & Ph & $\propto \nu^{-1}$ & Yes & Yes (few seconds) \\
Ionosphere - 2sn ord. (Faraday rotation) & Rotation & Both & $\propto \nu^{-2}$ & Yes & Yes (few seconds) \\
Ionosphere - 3rd ord. & Scalar & Ph & $\propto \nu^{-3}$ & Yes & Yes (few seconds) \\
Ionosphere - scintillations & Diagonal & Amp & -- & Yes & Yes (few seconds) \\
Dipole beam & Full-Jones & Both & -- & Yes & Yes (minutes) \\
Bandpass & Diagonal & Amp & -- & No & No \\
\end{tabular}
\begin{tablenotes}
    \item[a] In linear polarisation basis.
    \item[b] The matrix affects phases, amplitude or both.
\end{tablenotes}
\end{threeparttable}
\caption{Type of systematic effects we isolated in LOFAR data. For each effect we describe the associated Jones matrix, the frequency dependency and if it is time or direction dependent.}\label{tab:effects}
\end{table*}

\subsection{Clock}
\label{sec:clock}

The LOFAR stations are equipped with a GPS-corrected rubidium clock. All CSs are connected to the same clock, while each RS and IS has a separate clock. The timestamps made by the clocks can drift by up to 20 ns per 20 min, which corresponds to about a radian per minute at 150 MHz. Clocks are periodically re-aligned using GPS signals. This creates a time-dependent delay between all CSs (assumed as reference) and any other station. Since the same clock is used for both polarisations, the effect is represented by a scalar. Clock errors are equivalent to time delays, therefore their effect is proportional to frequency. The effect can be more severe than the ionospheric corruptions in the HBA frequency range.

\subsection{Polarisation alignment}
\label{sec:polalign}

In a LOFAR station, the two data streams from the X and Y polarisations are formed independently and different station calibration tables are applied to the two data streams. A station calibration is an automatic procedure that compensates for different delays and sensitivity of the individual dipoles within a station. Station calibration tables can imprint an artificial constant delay offset between the two data streams. This offset is constant in time and can be described as a phase matrix with either only the XX or the YY term $\neq 0$. Since this effect is a phase-only effect, one station is taken as reference and their streams are considered synchronised.

\subsection{Ionosphere}
\label{sec:ionosphere}

The ionosphere is a layer of partially ionised plasma surrounding the upper part of the atmosphere of the Earth. The peak of the free electron density lies at a height of $\sim300$~km but the ionosphere extends, approximately, from 75 to 1000 km. The ionosphere is a major source of systematic corruptions in LOFAR observations. A full treatment of the effect of the ionosphere on LOFAR observations is given in \cite{deGasperin2018a}. In this section we summarise part of that paper. The major effect of the ionosphere on interferometric observations is the introduction of a time- and direction-dependent propagation delay \citep[e.g.][]{Intema2009}. The effect is caused by a varying refractive index $n$ of the ionospheric plasma along the wave trajectories. The total propagation delay, integrated along the line of sight (LoS) at frequency $\nu$, results in a phase rotation given by

\begin{equation}\label{eq:delay}
 \Phi_{\rm ion} = - \frac{2\pi\nu}{c} \int_{\rm LoS} \left( n - 1 \right)\ {\rm d}l.
\end{equation}

For signals with frequencies higher than the ionospheric plasma frequency $\nu_p \simeq 10$~MHz, the refractive index $n$ can be expanded \citep[see e.g.][]{Datta-Barua2008} in powers of inverse frequency. The first order term ($\Phi_{\rm ion}\propto \nu^{-1}$) depends only on the density of free electrons integrated along the LoS, also called total electron content (TEC). The associated Jones matrix is a scalar as the effect corrupts both X and Y polarisation signals in the same way. This is the dominant term; for most radio-astronomical applications at frequencies higher than a few hundred Megahertz, higher order terms can be ignored. The second order term ($\Phi_{\rm ion}\propto \nu^{-2}$) causes Faraday rotation. This term depends on TEC and the magnetic field of the Earth. In the linear polarisation basis, it can be described by a rotation matrix. We note that a rotation matrix with such a fixed frequency dependency has only one degree of freedom (per time slot and direction). The third order term ($\Phi_{\rm ion}\propto \nu^{-3}$) is usually ignored but can become relevant for observations at frequencies below 40~MHz \citep{deGasperin2018a}. This term depends on the spatial distribution of the electrons in the ionosphere \citep{Hoque2008}; it becomes larger if electrons are concentrated in thin layers and not uniformly distributed. The third order ionospheric effect is also a scalar. For widely separated stations all ionospheric effects vary on a timescale of seconds. Because of their dependence on local ionospheric conditions, all ionosphere-related terms are direction dependent.

\subsection{Beam}
\label{sec:beam}

In this section, ``beam'' refers to the dipole beam. This is the common beam shape that each dipole in a LOFAR station has and it is fixed in the local horizontal coordinate system. The X and Y dipoles have a very different response, therefore the LOFAR (dipole) beam representation is a $2\times2$ full-Jones matrix. Since in this paper we are dealing with calibrator sources located at the phase centre, the array beam, i.e. the beam response of the whole station, is essentially constant over time and equal to 1; therefore the array beam is ignored. This is an approximation; some effects are currently not modelled in the LOFAR beam libraries and can contribute to small variations of the array beam in time. Two examples of this are the HBA analogue beam former or the mutual coupling of LBA dipoles. However, at the phase centre these effects are expected to be secondary. The beam matrix is time dependent as the direction of the source changes along the observation. The beam response is maximal if the source is located at the zenith and low if the source is close to the horizon. This matrix is estimated using an analytical model of the dipole response \citep{VanHaarlem2013}.

\subsection{Bandpass}
\label{sec:bandpass}

The LOFAR bandpass is shaped by a combination of effects. In the LBA case, the main effect is the frequency dependency of the dipole beam that has a peak efficiency near the resonance frequency of the dipole. In the HBA, a $\sim 1$~MHz ripple across the band comes from standing waves in the cables connecting the tiles with the electronics. In both antenna systems, a smaller effect ($\sim 0.1\%$) comes from the improper removal of the correlator conversion to frequency domain through a poly-phase filter. This process leaves a frequency-dependent signature in the data that is partially corrected within each 0.2 MHz-wide sub-band (SB) at correlation time. This effect is still visible at very high frequency resolution (3.052 kHz).

Since the time dependency of the beam is discussed in the previous section, the bandpass is effectively a time-independent effect, which affects the visibility amplitudes in the same way for both polarisations. As a consequence, the LOFAR bandpass Jones matrix is expected to be a real scalar value. However, the unmodelled differences among X and Y dipoles create small deviations from this ideal case. We therefore treat the bandpass as a diagonal Jones matrix.

\section{Observations}
\label{sec:observations}

\begin{table}
\centering
\begin{threeparttable}
\begin{tabular}{lcc}
Target calibrator & \multicolumn{2}{c}{3C\,196} \\
Antenna & LBA & HBA \\
\hline
Project code & LC5\_017 & LC3\_028 \\
Observation date & 05-06 Feb 2016 & 26-27 Feb 2015 \\
Integration time & 8 hrs & 6 hrs \\
Total timestamps & 7200 & 5400 \\
Time resolution & 1.0 s & 2.0 s \\
 -- after averaging & 4.0 s & 4.0 s \\

Average freq.  & 54.1 MHz & 151.6 MHz \\
Frequency range & 42--66 MHz & 115--189 MHz \\
Bandwidth (fract.) & 23.8 MHz (44\%) & 74.8 MHz (49\%) \\ 
Total channels & 488 & 381 \\
Freq. resolution & 3.052 kHz & 3.052 kHz \\
 -- after averaging & 48.828 kHz & 195.313 KHz \\

Stations (baselines) & 35 (595) & 58 (1653) \\
Station mode & LBA\_OUTER & HBA\_DUAL\_INNER \\
\end{tabular}
\end{threeparttable}
\caption{Parameters of LOFAR LBA and HBA observations.}\label{tab:obs}
\end{table}

Radio calibrators are significantly unresolved, bright sources that dominate the integrated flux of the surrounding field. Observations pointed at such sources are used to obtain a sensible calibration of LOFAR data. At the frequency and resolution of LOFAR, only a handful of sources meet these requirements \citep{Scaife2012}. Among those, we have shown that 3C\,196, 3C\,295, and 3C\,380 are good calibrators for LOFAR LBA. However, owing to its extended component on scales $\sim20\arcsec$, 3C\,380 shows some decrease in the flux density in all baselines that include the most RSs. All mentioned calibrators have flux densities $\gtrsim 100$~Jy at 100~MHz and only 3C\,295 has a turnover that might affect the calibration of the lowest frequencies ($<40$ MHz). For LOFAR HBA the major limitation is the model precision and flux concentration of the source at $5\arcsec$ resolution. In this case, good calibrators are: 3C\,196, 3C\,295, 3C\,48, and to lesser extent, 3C\,147. Fortunately, almost always one of these sources is at a high enough elevation ($>30 \deg$) to be used as calibrator. 

For this analysis we used archival LOFAR LBA and HBA observations pointed at 3C\,196. The LBA observation was performed on February 5 to 6, 2016 using the LBA\_OUTER mode. This mode uses the outermost 48 dipoles of each LBA station. It provides a station width of 81~m, which translates in a primary beam full width at half maximum (FWHM) $\sim 4\deg$ at 54 MHz. The HBA observation was performed on February 26 to 27, 2015 using the HBA\_DUAL\_INNER mode. For the HBA systems, the dipoles of CSs are divided into two substations. These substations have a larger field of view (FoV; with FWHM $\sim 4\deg$) than the RSs. To harmonise the FoV, in this observing mode all RSs have a reduced collecting area that matches the one of the core substations. The most important parameters for the observations are listed in Table~\ref{tab:obs}.

In both cases the telescope was configured to observe both polarisations and to produce four correlation products per baseline. For the LBA observation the frequency coverage was 42--66 MHz (bandwidth: 23.8 MHz). The frequency band was divided into 122 SBs, each 195.3125~kHz wide. Each SB is then subdivided into 64 channels of 3.052~kHz. The time resolution was set to 1~s. These high frequency and time resolution parameters were chosen to have a better handle on radio frequency interference (RFI) detection, to surgically exclude fast and narrow-band RFI without losing useful data. The HBA observation was performed with similar parameters. In this case, the frequency coverage was set to 115--189 MHz (bandwidth: 74.8 MHz).

For the LBA observation we used 24 CSs and 13 RSs, all located within the Dutch border. This provided a baseline range between 60~m and 84~km. Because of technical malfunctions two stations were excluded at the beginning of the calibration. For the HBA observation we used 48 sub-CSs and 13 RSs. Three of the CSs were removed as a result of malfunction.

In what follows we take the LBA observation as a practical example. However, the HBA procedure is very similar and each LBA solution plot shown has its HBA counterpart displayed in Appendix~\ref{sec:HBA}. An important difference between LBA and HBA is that the station beam of the latter has an intermediate analogue beam-forming step (tile beam) that prevents LOFAR HBA from observing in multiple arbitrary directions at the same time. Therefore, while LBA solutions obtained on a calibrator field can be applied to any simultaneous target beam in real time, for HBA the beam has to move from the calibrator to the target field, assuming the latter is not within the tile beam (FWHM $\sim 20$\deg). This implies an extrapolation in time of any time-dependent systematic effect that one wants to transfer (e.g. the station clock drift).

\section{Calibration strategy}
\label{sec:strategy}

\begin{figure}
\centering
 \includegraphics[width=0.5\textwidth]{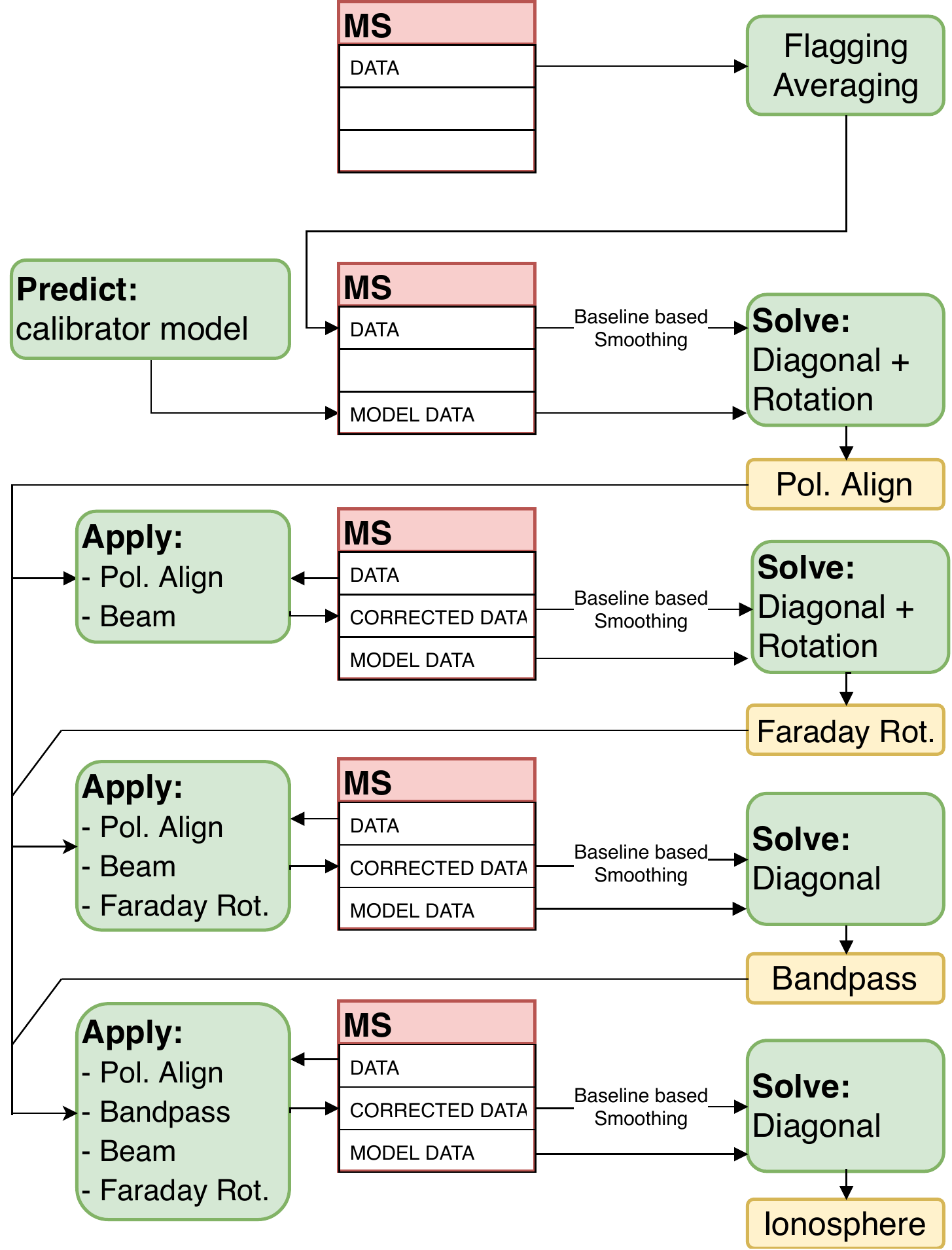}
 \caption{Schematic view of the steps used to calibrate LBA and HBA data for bright, compact point sources. Steps indicated in green are solve, apply, and predict steps and are carried out with DPPP \citep[see Appendix~\ref{sec:dppp};][]{VanDiepen2018}. Steps shown in yellow consist of solutions manipulations and are carried out by LoSoTo (see Appendix~\ref{sec:losoto}). Each solve step has an input data column and also uses data from the model. Each apply step has an input data column and an output data column. In each apply step all the listed calibration tables are applied in the specified order.}
 \label{fig:calibscheme}
\end{figure}

The calibrator data reduction pipeline consists of a number of steps outlined in Fig.~\ref{fig:calibscheme}. In the image, data sets are represented by the red boxes, sets of visibilities are listed into the box. Predicting visibilities, manipulating the data, finding station-based solutions and applying these to the data is done with the Default Preprocessing Pipeline (DPPP; green steps, see Appendix~\ref{sec:dppp}). Solve steps ignore baselines shorter than $300 \lambda$, which prevents the unmodelled large-scale mission from the Galaxy from biasing our results. Since we are working on a calibrator field, the S/N is high enough that we can solve  on a single time step and frequency channel. This allows for easy parallelisation of the code by working on each channel simultaneously as independent time streams. On the other hand, the estimation of the few parameters describing the systematic effects must be carried out by combining all frequency channels and, in some cases, polarisations. This is always done in solution space by a separate software called the LOFAR Solution Tool (LoSoTo; yellow steps, see Appendix~\ref{sec:losoto}). The aim of the whole process is to isolate the systematic effects that are direction independent and can therefore be transferred to the target field.

\begin{figure*}
\centering
 \includegraphics[width=\textwidth]{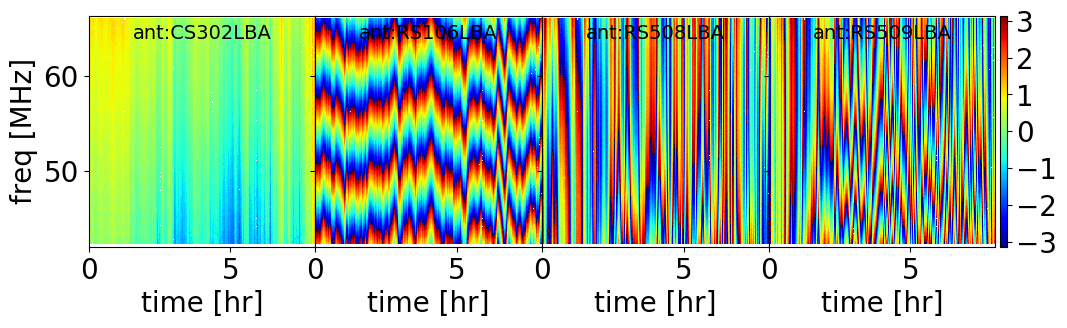}\\
  \includegraphics[width=\textwidth]{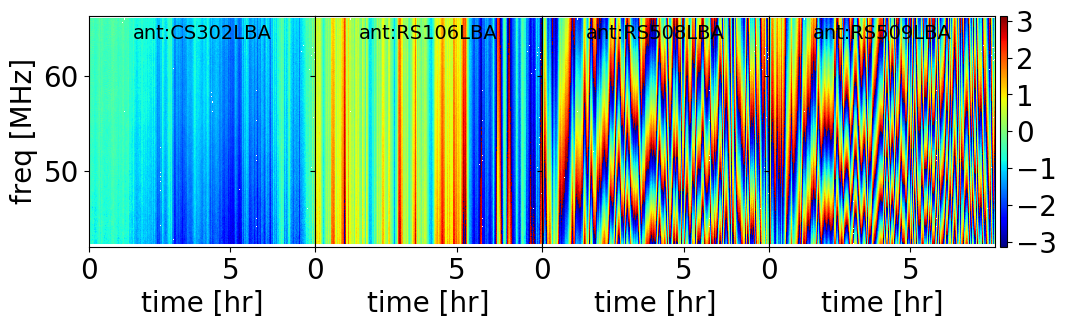}\\
 \includegraphics[width=\textwidth]{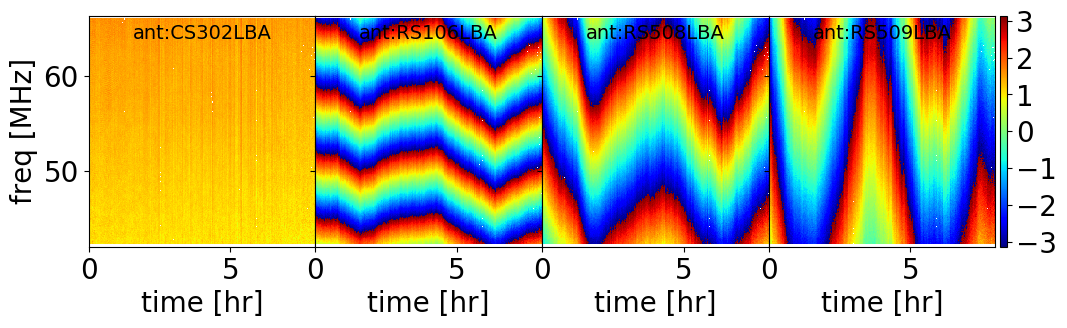}\\
 \includegraphics[width=\textwidth]{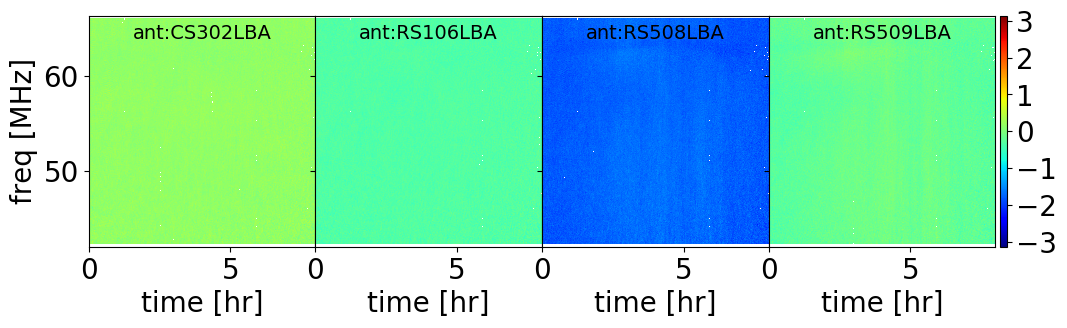}
 \caption{Phase solutions in radians for four different stations (CS302, RS106, RS508, and RS509) plotted as a function of observing time (x-axis) and frequency (y-axis). Colour goes from $-\pi$ (blue) to $+\pi$ (red). All phases are referenced to station CS002, at the array centre. \textit{First panel:} Phase solutions for the XX element of a diagonal Jones matrix obtained at the beginning of the calibration. Those solutions encode all the systematic effects affecting LOFAR LBA phases. \textit{Second panel:} Same as above but after the subtraction of the clock systematic effect,  only the ionosphere is visible. \textit{Third panel:} Same as the first panel, but after the subtraction of the ionospheric systematic effect, only the clock is visible (CS302 has the same clock of the reference). \textit{Bottom panel:} Same as above but solving after the subtraction of all recovered systematic effects, i.e. at the end of the calibration pipeline. The uniformity of the plots shows we are able to remove systematic effects with high accuracy. The HBA equivalent is shown in Fig.~\ref{fig:ph_HBA}.}
 \label{fig:ph_LBA}
\end{figure*}

\begin{figure*}
\centering
 \includegraphics[width=\textwidth]{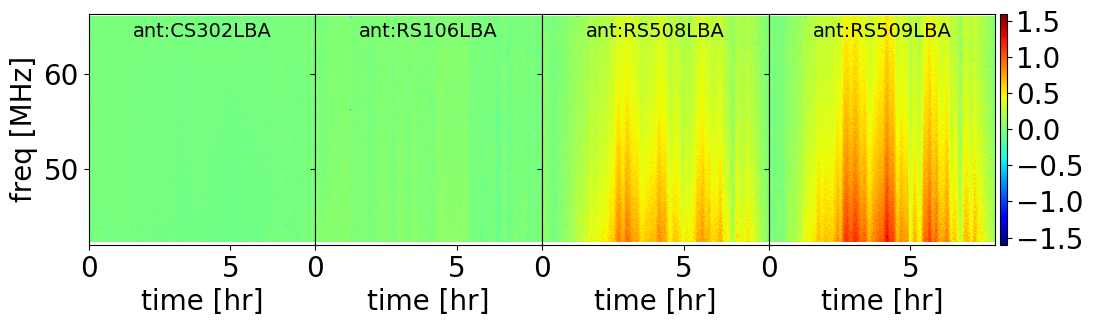}
 \caption{Same as Fig.~\ref{fig:ph_LBA} but is the rotation angle in radians of the Jones rotation matrix is colour coded. The two most RSs show clear evidence of Faraday rotation. The two stations are close in location, therefore the effect is similar. The HBA equivalent is shown in Fig.~\ref{fig:rot_HBA}.}
 \label{fig:rot_LBA}
\end{figure*}

The LoSoTo software can generate plots as those shown in Figs.~\ref{fig:ph_LBA} and \ref{fig:rot_LBA}. We present the phase solutions for four LBA stations at the beginning of the calibration process. The solutions are obtained by solving simultaneously for a diagonal Jones matrix and a rotation Jones matrix as described in Appendix~\ref{sec:dppp}. In this way, effects that can be described by a rotation matrix (e.g. Faraday rotation) are isolated from other effects. Fig.~\ref{fig:ph_LBA} shows the first element (i.e. the solutions relative to the XX polarisation product) of the diagonal matrix before and after the subtraction of all known systematic effects recovered during the calibration procedure. As evident from the uniformity of the plot we were able to isolate the majority of the systematic effects with high accuracy.

\subsection{Preparation}

The first steps, performed immediately following the observation, include the flagging of the RFI with AOflagger \citep{Offringa2012} and the subsequent averaging of the data to a manageable size. The next step is further flagging of known problematic antennas and of periods in time when the calibrator field is below 20\deg{} elevation, where the dipole response is highly suppressed. A final averaging step is performed to bring data to 4 s time resolution and four channels per SB (195.3 KHz) frequency resolution. Given the lower impact of the ionosphere at higher frequencies, for the HBA data sets the frequency averaging can be increased to one channel per SB. 

In order to save computing time, the calibrator visibilities are predicted from a calibrator model. This process is performed only once at the beginning of the pipeline. We use a model of 3C\,196 described by four Gaussian components, in which each component has a spectrum described by  a second order log-polynomial (V.N. Pandey priv. comm.)

Before any solve step we perform a baseline-based smoothing to exploit the time coherency of all systematic effects. This is accomplished by smoothing the data along the time axis with a running Gaussian independently for each channel and polarisation. The timescales over which the ionospheric effects can be considered to be coherent (i.e. with negligible phase changes), and thus over which we can safely smooth, depend on the distance between the two stations that form a baseline and on the inverse of the frequency (to first order). The distance dependence arises owing to the turbulent nature of the ionosphere and the standard deviation of the phase difference between two stations scales with their separation, $r$, as $r^{\beta/2}$ with $1.5 \lesssim \beta \lesssim 2$ \citep{Mevius2016}. Therefore, we adopt a frequency- and baseline-dependent scaling for the width in time of the smoothing Gaussian of: FWHM $\propto \nu r^{-1/2}$, where $\beta = 1$ was chosen for simplicity (in general, $\beta$ is time dependent). This scaling is normalised to prevent over-smoothing at the lowest frequencies and longest baselines on the relevant timescales ($\sim 5-10$ sec for typical ionospheric conditions). Flagged data are ignored in the process. Owing to the Gaussian smoothing, the variance $\sigma_0^2$ of the data is expected to be reduced to
\begin{equation}
    \sigma_f^2 \approx \frac{\sigma_0^2}{\sqrt{2 N \sqrt{\pi}}},
\end{equation}
where N is the standard deviation of the filter Gaussian and varies between 1 and 20 depending on baseline length. Therefore, we expect a reduction in terms of noise in the data that ranges from 2 to 10 going from longest to shortest baselines.


\subsection{Polarisation alignment}

\begin{figure*}
\centering
 \includegraphics[width=\textwidth]{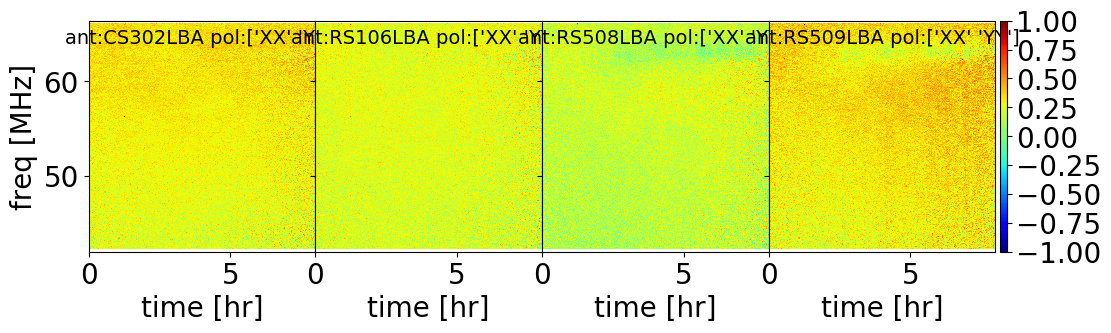}\\
 \includegraphics[width=\textwidth]{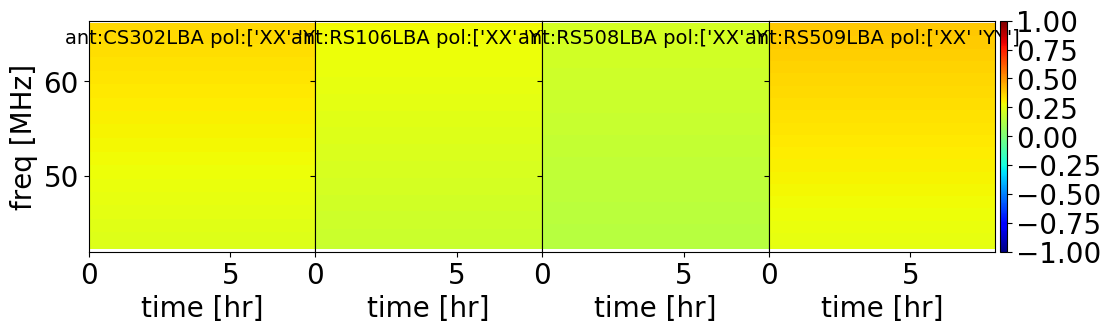}\\
 \includegraphics[width=\textwidth]{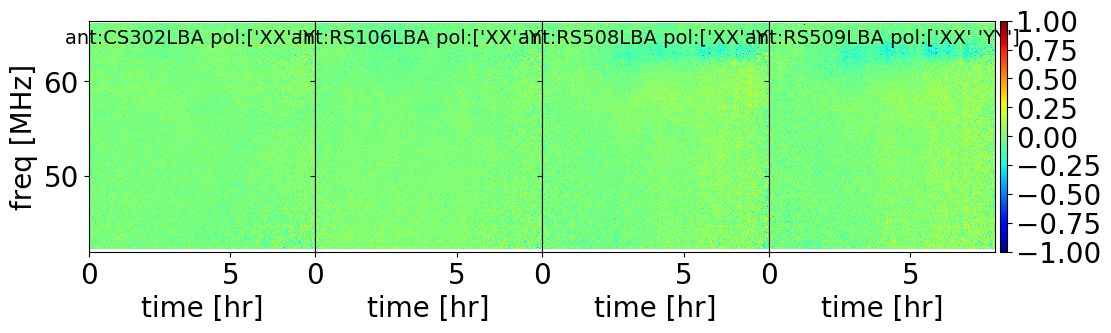}
 \caption{Same as Fig.~\ref{fig:ph_LBA}. \textit{Top panel:} Differential (XX-YY) phase solutions. \textit{Mid-panel:} The time-independent delay fit performed by LoSoTo. \textit{Bottom panel:} Differential phase solutions after the subtraction of the delay. The HBA equivalent is shown in Fig.~\ref{fig:pa_HBA}.}
 \label{fig:pa_LBA}
\end{figure*}

During the first solve step we estimate values of a diagonal plus a rotation Jones matrix. The rotation matrix is included to capture the effect of Faraday rotation, the only rotation matrix identified in Table~\ref{tab:effects}. Any other effect ends up in the diagonal matrix. The first systematic effect we want to correct for will be the last along the signal path. This is a polarisation misalignment introduced by the station calibration tables. This has the form of a delay, therefore affects phases with a linear frequency dependency. To visualise the effect, the phases of one term of the diagonal solution matrix are subtracted from the phases of the other (XX -- YY). In observations of unpolarised sources, the result should be zero. However, LOFAR data show a misalignment visible at all frequencies (top panel of Fig.~\ref{fig:pa_LBA}). Using LoSoTo we fit a time-independent delay term across the entire bandwidth (second panel of Fig.~\ref{fig:pa_LBA}). This a good example to show how the degrees of freedom are strongly reduced from a very large number of solutions to just one number per antenna, i.e. the delay. This delay is instrumental and time independent. Therefore, it can be easily transferred to the target field(s).

\subsection{Faraday rotation}

The second step is the estimation of the Faraday rotation. Firstly, we align the polarisation data streams using the result of the previous section. Secondly, we use the theoretical dipole-beam estimation to correct for its effect. The correction of the dipole beam must be applied after the polarisation alignment as that corruption happens earlier in the signal path and the two matrices do not commute. The dipole beam does not compromise the estimation of the polarisation delay because the former is mostly an amplitude effect, while the polarisation delay is estimated using phases. After that, to avoid any possible leakage of the beam effect into the rotation matrix, we solve again for a diagonal plus a rotation Jones matrix. The rotation matrix should now contain only the Faraday rotation. Then, we use the solutions of the rotation matrix to estimate the time-dependent Faraday rotation by fitting a $\propto \nu^{-2}$ frequency dependency in solution space; see Fig.~\ref{fig:rot_LBA}. The estimated time stream of the differential Faraday rotation is shown in Fig.~\ref{fig:syseff_LBA}.

\subsection{Amplitude calibration}

\begin{figure*}
\centering
 \includegraphics[width=\textwidth]{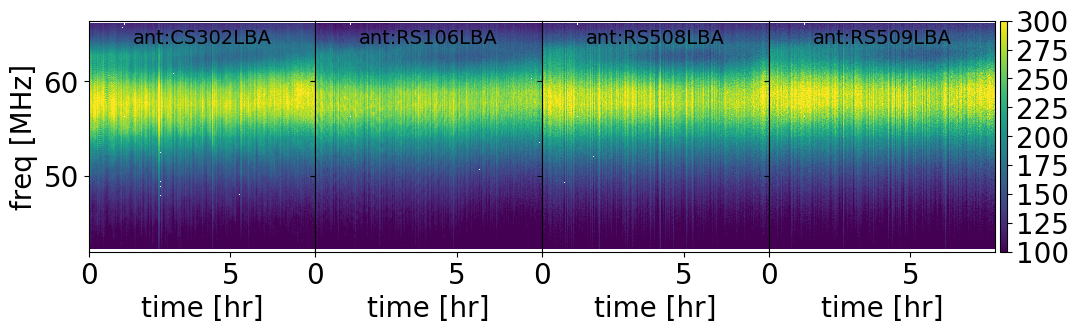}\\
 \includegraphics[width=\textwidth]{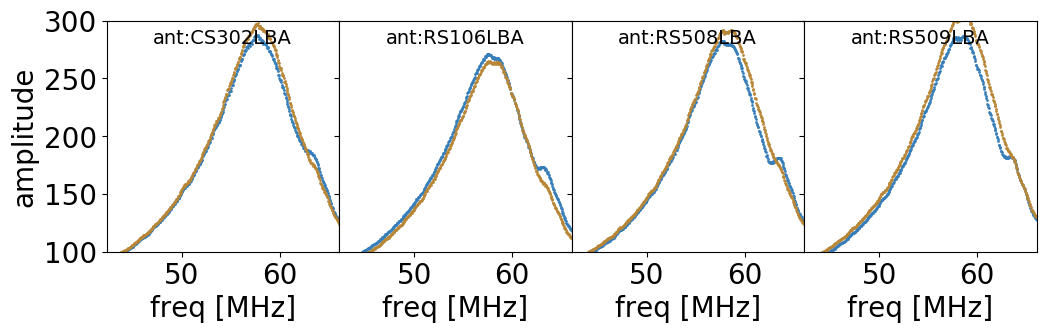}\\
 \includegraphics[width=\textwidth]{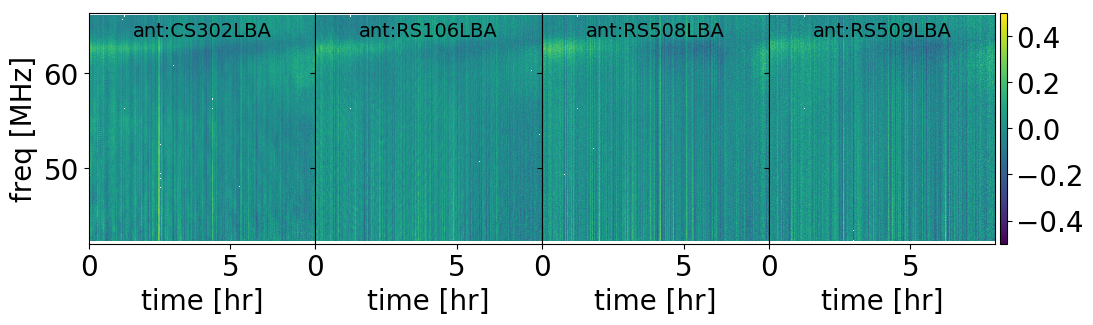}
 \caption{Same as Fig.~\ref{fig:ph_LBA}. \textit{Top panel:} Amplitude solutions for the XX element of the diagonal matrix. \textit{Mid-panel:} Time median of the amplitude solutions; this represents the instrument bandpass. Blue is for the XX polarisation and brown for the YY. \textit{Bottom panel:} Residuals after dividing the top panel by the time-independent bandpass. The HBA equivalent is shown in Fig.~\ref{fig:amp_HBA}}.
 \label{fig:amp_LBA}
\end{figure*}

The next step is to isolate the amplitude bandpass. After applying the polarisation alignment, the dipole beam and the Faraday rotation (in this order) we solve for a diagonal Jones matrix. We show the amplitude part of this matrix in Fig.~\ref{fig:amp_LBA}. Two major effects are present here: the bandpass itself, which is time independent, and the ionospheric scintillation that varies with time. To isolate the first we extract the median of each channel along the time interval spanning the entire observation; this produces the time-independent and direction-independent bandpass (second panel Fig.~\ref{fig:amp_LBA}) that can be exported to the target field(s). Assuming that our calibrator model is correct and that the dipole beam is accurate, this matrix takes care of re-scaling the flux density of the target(s) to the correct value. While in the ideal case the bandpass of the X and Y dipole should be the same, we keep these dipoles separate to compensate for beam model inaccuracies (unfortunately this doubles the degrees of freedom). The notch in the XX polarisation is likely due to the edge of the dipole wire, which is a loop. The size of the loop can vary from dipole to dipole and in certain cases can be wet, modifying the dipole theoretical response. The effect can appear on none, one, or both polarisations depending on conditions and it is currently under investigation (M.J. Norden priv. comm.). In the last panel of Fig.~\ref{fig:amp_LBA} we show the residuals after the bandpass subtraction. The series of thin, vertical lines are ionospheric amplitude scintillations, while the slow variations in time are inaccuracies in the dipole beam model.

\subsection{Clock and ionospheric calibration}

\begin{figure*}
\centering
 \includegraphics[width=0.31\textwidth]{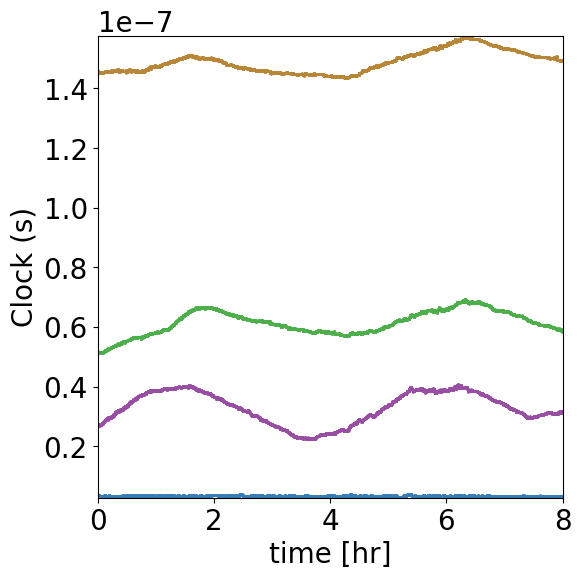}
 \includegraphics[width=0.33\textwidth]{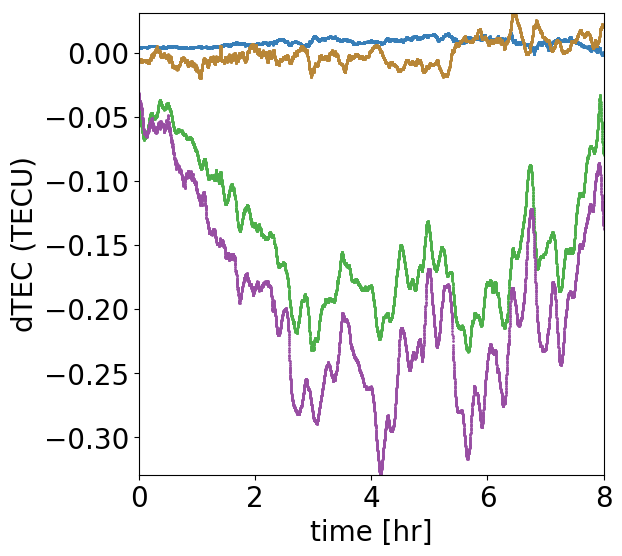}
 \includegraphics[width=0.34\textwidth]{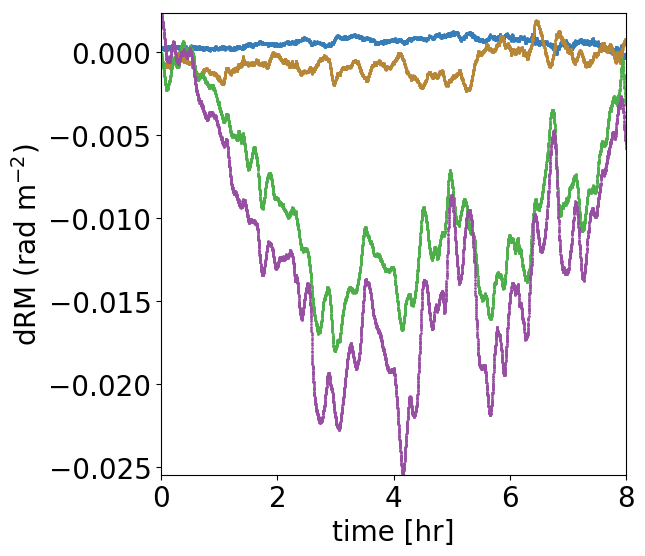}
 \caption{Ionospheric systematic effects affecting phases for the same four stations of Fig.~\ref{fig:pa_LBA}. From top to bottom in the first panel: RS106 (brown), RS508 (green), RS509 (purple), and CS302 (blue). From left to right: clock delays, first order ionospheric delay, and Faraday rotation. The CS has uniformly zero clock delays as its clock is the same as the reference station (CS002). RS508 and RS509 TEC values track each other as the two stations are relatively close by. The TEC unit (TECU) is defined as $10^{16}$~m$^{-2}$, which is the order of magnitude typically observed at zenith during the night. The clear correlation between dTEC and dRM (second and third panel) is because differential Faraday rotation is to a large extent caused by the difference in integrated TEC multiplied with the parallel magnetic field. The HBA equivalent is shown in Fig.~\ref{fig:syseff_HBA}.}
 \label{fig:syseff_LBA}
\end{figure*}

The final step is the calibration of clock and ionospheric delays. We pre-apply all previously found solutions and solve again for a diagonal Jones matrix. While these delays are scalars, we solve for a diagonal matrix to keep track of the residual differences in the X and Y data streams. The phase solutions obtained in this way are a combination of two effects: clock and ionosphere (first order). These effects have a different frequency dependency of $\propto\nu$ and $\propto\nu^{-1}$, respectively. We now apply a procedure called clock/TEC separation \citep[see e.g.][]{vanWeeren2016b,deGasperin2018a} to find the best fit of these two parameters across the bandwidth for each time step. The outcome of this process is shown in the first two panels of Fig.~\ref{fig:syseff_LBA}. Although the clock/TEC separation is done independently for each time step, the clock drifts and ionospheric TEC values show a clear temporal correlation. The clock drifts also present the typical segmented shape due to instant GPS corrections that happen regularly to prevent the clock from drifting too much. Since the clock delay is a direction-independent effect, the solutions can be transferred to the target field(s). All phase-derived values are differential with respect to CS002, as a consequence the derived ionospheric effects are smaller for stations close to the LOFAR core. The TEC values of RS508 and RS509 track each other because of the proximity of the two stations, i.e. their beams see through a similar ionosphere. For observations that go below 40~MHz the third order ionospheric effect becomes non-negligible. In those cases the clock/TEC separation process must include another parameter to capture the $\nu^{-3}$ dependency of the term \citep{deGasperin2018a}.

\section{Image analysis}
\label{sec:images}

\begin{figure*}
\centering
 \includegraphics[width=\textwidth]{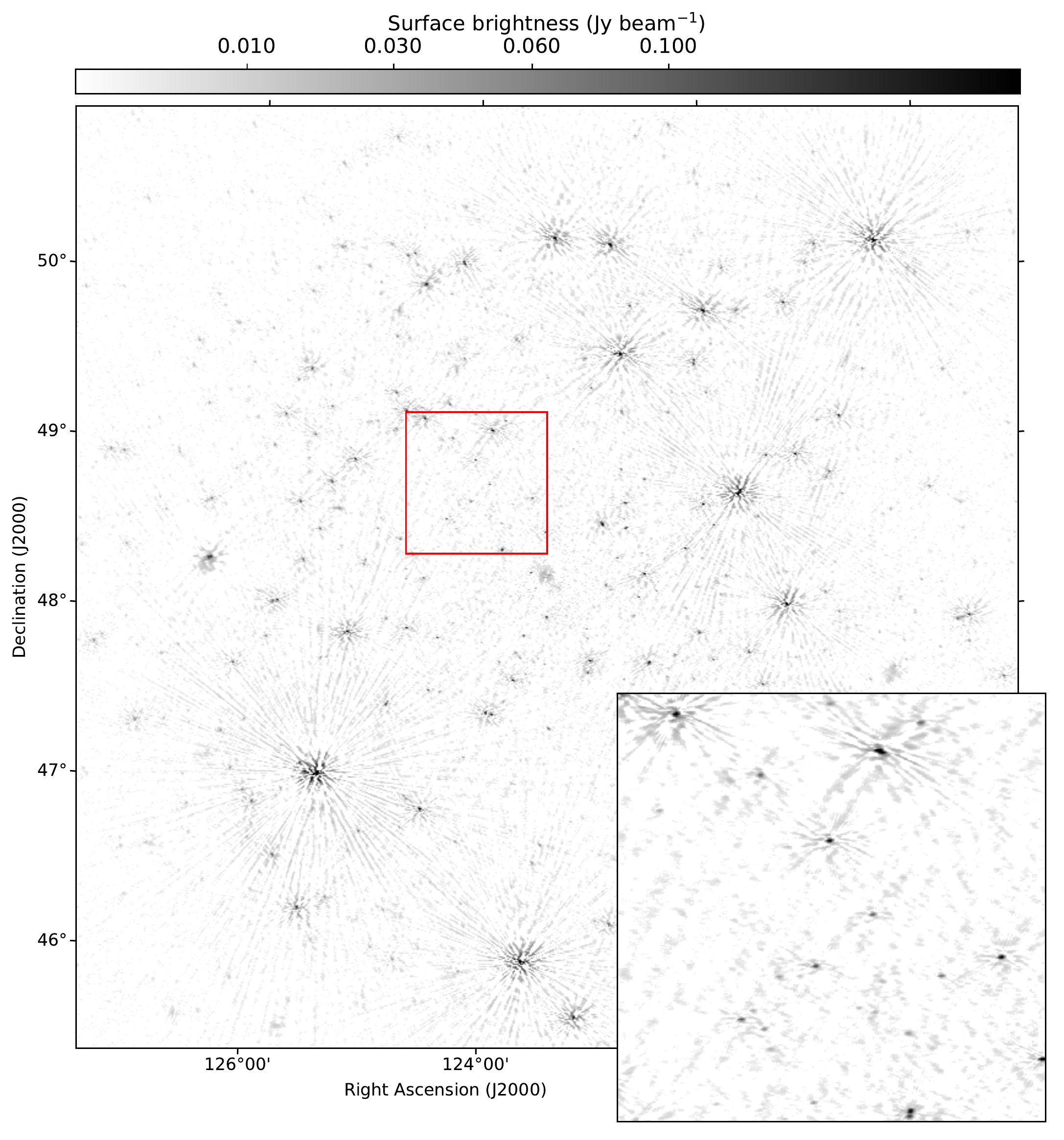}
 \caption{Image of the field around 3C\,196 (subtracted) at 54 MHz obtained with LOFAR LBA. The image resolution is \beam{26}{14} and the rms noise is 3~\mjybeam. The red square shows the region that is zoomed-in in the bottom right corner.}
 \label{fig:map_LBA}
\end{figure*}

\begin{figure*}
\centering
 \includegraphics[width=\textwidth]{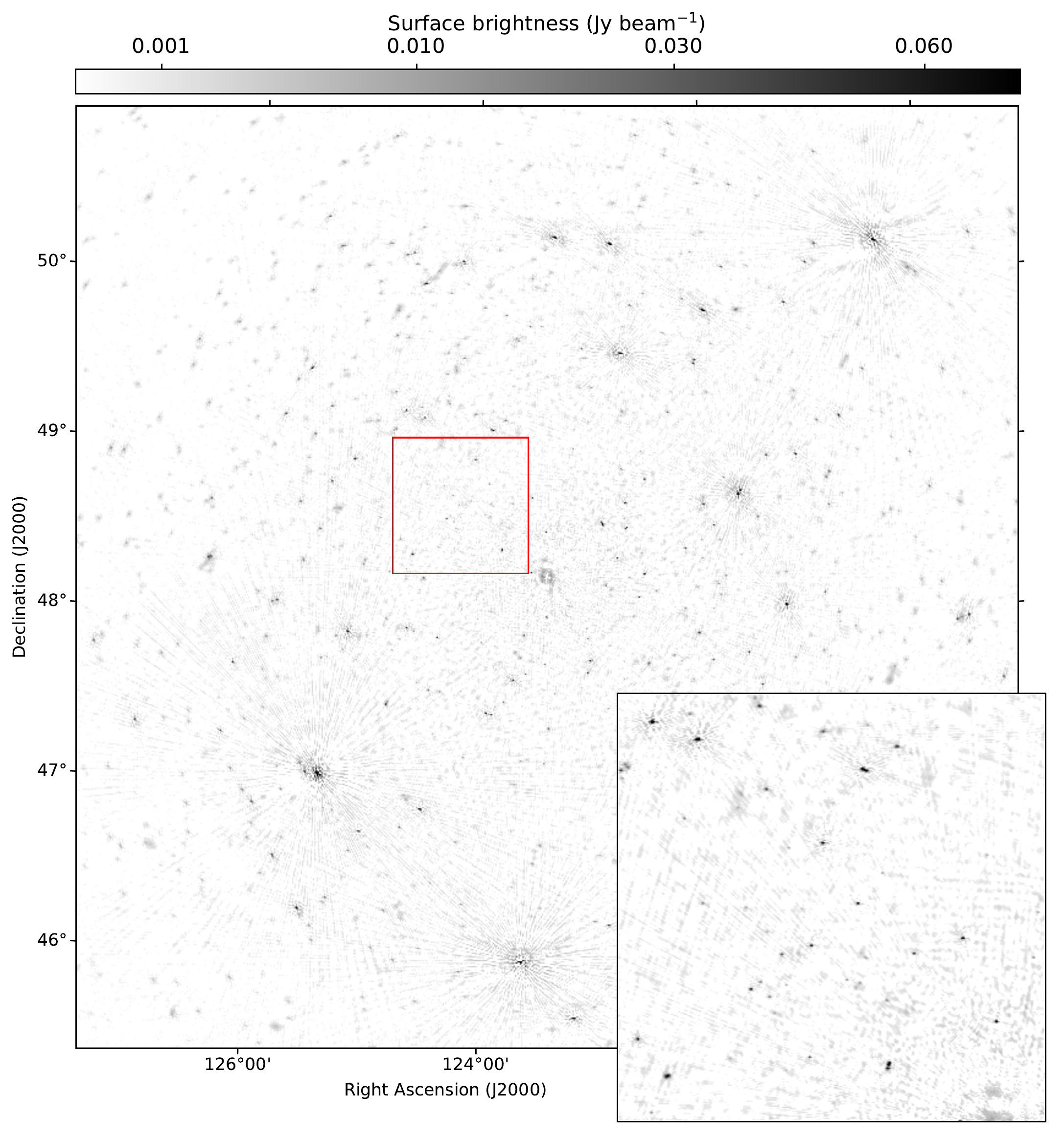}
 \caption{Image of the field around 3C\,196 (subtracted) at 152 MHz obtained with LOFAR HBA.  The image resolution is \beam{16}{10} and the rms noise is 450~\mujybeam. The red square shows the region that is zoomed-in in the bottom right corner.}
 \label{fig:map_HBA}
\end{figure*}

\begin{figure*}[ht!]
\centering
 \includegraphics[width=\textwidth]{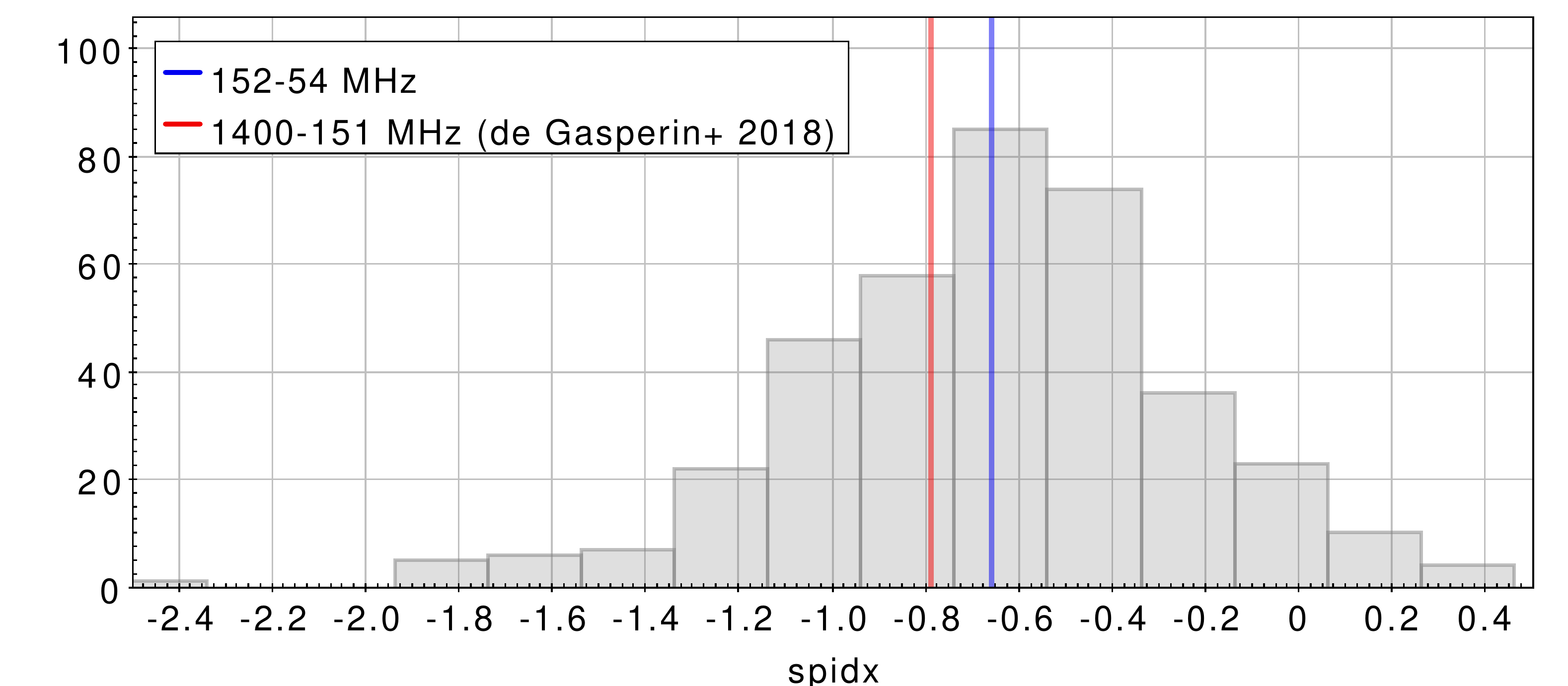}
 \caption{Spectral index distribution of the 520 matched sources from Fig.~\ref{fig:map_LBA} and Fig.~\ref{fig:map_HBA}. The mean of the distribution is $\alpha_{54}^{152}=-0.66$ and the median is $\alpha_{54}^{152}=-0.63$. The mean spectral index found cross-matching half a million sources from TGSS (151 MHz) and NVSS (1400 MHz) surveys is also shown \citep[$\alpha_{151}^{1400} = -0.79$;][]{deGasperin2018}.}
 \label{fig:spidx}
\end{figure*}

As a final verification we applied the solutions obtained in the last step to the data. Then, we subtracted the calibrator from the visibilities using the best available model to facilitate the imaging of the rest of the field. We note that all direction-dependent effects (ionospheric first and second order, amplitude scintillations, and dipole beam) are evaluated in the direction of the calibrator, which is the phase centre. We expect increasingly strong artefacts, mostly around bright sources, as we move away from the phase centre. The images of the field surrounding the calibrator 3C\,196 were produced with WSclean\footnote{LBA: $4000 \times 4000$ pixels of $5\arcsec \times 5\arcsec$ area. HBA: $5000 \times 5000$ pixels of $4\arcsec \times 4\arcsec$ area.} \citep{Offringa2014} and are shown in Figs.~\ref{fig:map_LBA} and \ref{fig:map_HBA} for LBA (54 MHz) and HBA (152 MHz), respectively. The average rms noise of the LBA image is $\sim 3$~\mjybeam. The expected rms noise is $\sigma = {\rm SEFD}\ \sqrt{N \left( N-1 \right) \Delta \nu \Delta t} \approx 1$~\mjybeam. This calculation uses ${\rm SEFD} \sim 26$~kJy  \citep[system equivalent flux density;][]{VanHaarlem2013} at 54 MHz, $N=35$ (number of stations), $\Delta\nu=23.8$~MHz, and $\Delta t=0.9 \times 8$~hrs (assuming 10\% of flagged data). The factor of 3 difference is likely due to the missing direction-dependent calibration. The average rms noise of the HBA image is $\sim 450$~\mujybeam. The same calculation but assuming ${\rm SEFD} \sim 3$~kJy at 152 MHz, $N=58$, $\Delta\nu=74.8$~MHz, and $\Delta t=0.9 \times 6$~hrs gives an expected rms noise of $\sigma \approx 50$~\mujybeam. In both cases the real noise is higher by roughly a factor of two due to the weighting scheme used at imaging time.

The full scientific exploitation of the images requires a direction-dependent correction that is not covered in this paper. However, we show the potential of having large FoV observations at both LBA and HBA frequencies by making a spectral index analysis of the detected sources. A flattening of the spectral index at low frequencies is expected because of (i) absorption at low frequencies, (ii) spectral ageing at high frequencies, and (iii) the necessary break down of the assumption that the energy distribution of cosmic-ray electrons in radio sources is an infinite power law towards low frequencies. Studies of low-frequency spectral indices are limited because of the difficulties in collecting a large number of sources at ultra-low frequencies ($<100$~MHz).

We ran the source extractor pyBDSF \citep[Python Blob Detection and Source Finder][]{Mohan2015} on both images after a primary beam correction. The software identifies islands of pixels above three times the local rms noise. Then, the code uses these islands to create sources by fitting and combining Gaussians centred on pixels above five times the local rms noise. We removed all sources whose flux in the island is larger than two times the flux in the source, which removed most of the sources surrounded by strong artefacts. We then cross-matched the resulting source catalogues with a matching radius of 26\arcsec, i.e. the major axis of the LBA data set point spread function. Finally, we estimated the spectral index\footnote{Spectral index $\alpha$ defined as $S_\nu \propto \nu^\alpha$, where $S_\nu$ is the source flux density.} for the 378 matched sources using the integrated flux density and finding the distribution shown in Fig.~\ref{fig:spidx}, with a mean $\alpha_{54}^{152}=-0.66$ and a median $\alpha_{54}^{152}=-0.63$. 

When working with spectral indexes a number of caveats need to be considered. Firstly, for a given frequency the dominant population of sources is different at different flux densities. At GHz frequency, a shallow survey mostly finds powerful lobes of FR\,II radio galaxies and some nearby FR\,I radio galaxy; in deeper observations AGN cores and star-forming galaxies would become the dominant populations \citep{Wilman2008}. All these populations have a different average spectral energy distribution behaviour that can bias the conclusions. Furthermore, the completeness of a spectral index catalogue depends on the depth of two (or more) surveys. Usually, surveys at higher frequencies are deeper (assuming a reasonable spectral index) than the low-frequency counterpart. This implies that a large number of faint flat-spectrum sources go undetected in the low-frequency survey and a smaller number of faint steep-spectrum sources also go undetected in the high-frequency survey. To obtain a reliable mean spectral index value we need to apply a cut at one of the two frequencies. In our case, the LBA image is shallower than the HBA image, such that even sources at the LBA detection limit with spectral index $\alpha_{54}^{152}=-2$ should have a high frequency counterpart detected within $5\sigma$ confidence level. In fact, we detected 88\% of LBA sources in the HBA image and the non-detections are not concentrated among the faintest sources. This implies that we are likely missing a (small) number of counterparts due to source mismatching or misclassification of artefacts. On the other hand only 46\% of HBA sources have an LBA counterpart. In order to compare our results with literature we need to apply a cut to the HBA data so that the majority of the sources have an LBA counterpart. Applying the cut $S_{\rm peak - 152 MHz} > 40$~mJy should provide an LBA counterpart for all sources with spectral index $>-2$. By applying this cut, we found an LBA counterpart for 90\% of HBA sources. The mean spectral index is now $\alpha_{54}^{152}=-0.50$ and the median $\alpha_{54}^{152}=-0.51$. Eddington bias can slightly overestimate this values. These values are higher (implying a flatter spectral energy distribution) than what is found in the literature for higher frequency ranges. For instance, in deep fields between 150 MHz and 1.4 GHz, the average values reported are $-0.87$ \citep{williams2013}, $-0.79$ \citep{intema2011}, $-0.78$ \citep{isw+10}, $-0.82$ \citep{sds+09}, and $-0.85$ \citep{ishwara2007}. On a larger sample from shallower survey data, the average spectral index in that frequency range is again $\alpha_{151}^{1400} = -0.79$ \citep{deGasperin2018}. Using LOFAR data at 150 MHz together with 1400 MHz information from surveys, \cite{Sabater2018} found a median spectral index of $\alpha_{150}^{1400} = -0.63$ for sources with $S_{\rm 1400 MHz} > 20$~mJy. Our results are in line with findings by \cite{VanWeeren2014} in which they found an average low-frequency spectral index $\alpha_{34}^{62}=-0.64$, which goes up to $\alpha_{34}^{62}=-0.5$ when inferred from source count scaling. All together, these results point towards a general flattening of the average spectral index of radio sources towards low frequencies.

\section{Conclusions}
\label{sec:conclusions}

In this paper we outlined a strategy to calibrate LOFAR LBA and HBA calibrator fields. The pipeline is implemented in a freely available code\footnote{\url{https://github.com/lofar-astron/prefactor}.}. The strategy relies on understanding the physics of all major systematic effects found in LOFAR data. We summarise these effects in Table~\ref{tab:effects}. Using physical priors, we are able to reduce the degrees of freedom of the calibration problem.

A full brute force calibration of the 8 hrs LBA data set presented here would require $\approx 1$ billion free parameters\footnote{4 (polarisations) $\times 8 \cdot 3600/4$ (time stamps) $\times 122*4$ (channels) $\times 35*2$ (stations amplitudes and phases) = 983 808 000}. With our procedures we demonstrate that the majority of the systematic effects can be represented by a significantly smaller number of free parameters: 35 (polarisation delays) $+ 30$k (bandpass) $+ 700$k (ionosphere and clock). As evident, fast ionospheric and clock variations are the dominant component. Comparable values, rescaled for the larger number of stations, are valid for HBA. Because of the inability of the HBA system to simultaneously observe an arbitrary target and a calibrator field, the LBA and HBA calibration procedures diverge after this initial step. Most importantly, in the HBA case some further corrections on the target field will be necessary to compensate for the extrapolated approximations of effects such as the clock drift. On the other hand, the higher S/N of HBA observations will make the target field direction-dependent calibration easier than for the LBA case.

As a final demonstration step we produced two images of the calibrator field at 54 and 152 MHz. The image at 54 MHz is currently the deepest image obtained at those frequencies reaching an rms noise of $\sim3$~\mjybeam{} with a resolution of \beam{26}{14} (expected thermal noise: $\sim1$~\mjybeam{}). In the HBA case we achieve an rms noise of $\sim 450$~\mujybeam{} with a resolution of \beam{16}{10} (expected thermal noise: $\sim50$~\mujybeam{}). We use these images to prove that the average spectral index values of radio sources tend to flatten at lower frequencies. 


\begin{acknowledgements}

FdG is supported by the VENI research programme with project number 639.041.542, which is financed by the Netherlands Organisation for Scientific Research (NWO).

AH acknowledges support by the BMBF Verbund-forschung under the grant 05A17STA.

RJvW acknowledges support from the ERC Advanced Investigator programme NewClusters 321271 and the VIDI research programme with project number 639.042.729, which is financed by the Netherlands Organisation for Scientific Research (NWO).

LKM acknowledges support from Oxford Hintze Centre for Astrophysical Surveys, which is funded through generous support from the Hintze Family Charitable Foundation. This publication arises from research partly funded by the John Fell Oxford University Press (OUP) Research Fund.

KLE acknowledges financial support from the Dutch Science Organization (NWO) through TOP grant 614.001.351.

JS is grateful for support from the UK STFC via grant ST/M001229/1.

WLW acknowledges support from the UK Science and Technology Facilities Council [ST/M001008/1]

LOFAR, the LOw Frequency ARray designed and constructed by ASTRON, has facilities in several countries, which are owned by various parties (each with their own funding sources) and are collectively operated by the International LOFAR Telescope (ILT) foundation under a joint scientific policy.

This research has made use of NASA's Astrophysics Data System.

\end{acknowledgements}


\bibliographystyle{aa}
\bibliography{papers-LBAsurvey}


\onecolumn
\begin{appendix}

\section{HBA images}
\label{sec:HBA}

\begin{figure*}[ht!]
\centering
 \includegraphics[width=\textwidth]{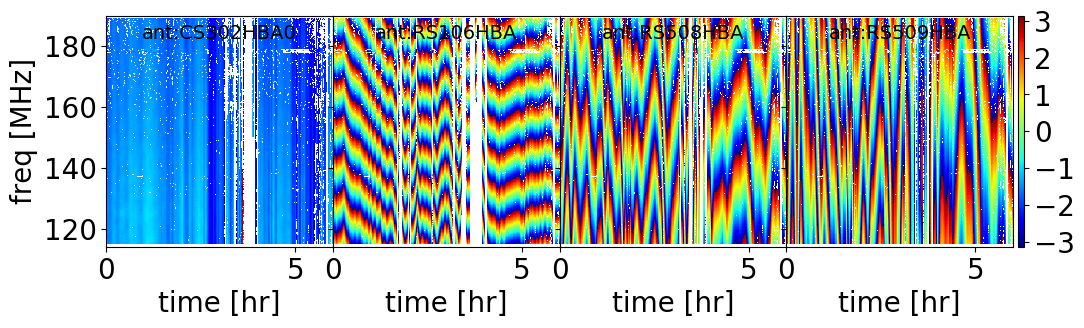}\\
 \includegraphics[width=\textwidth]{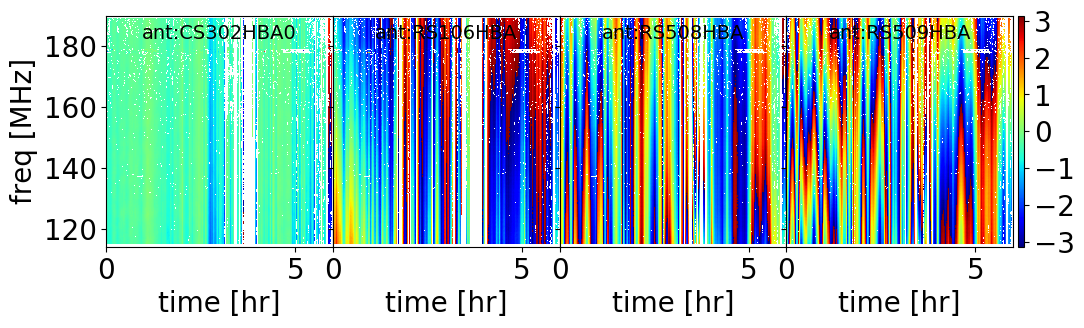}\\
 \includegraphics[width=\textwidth]{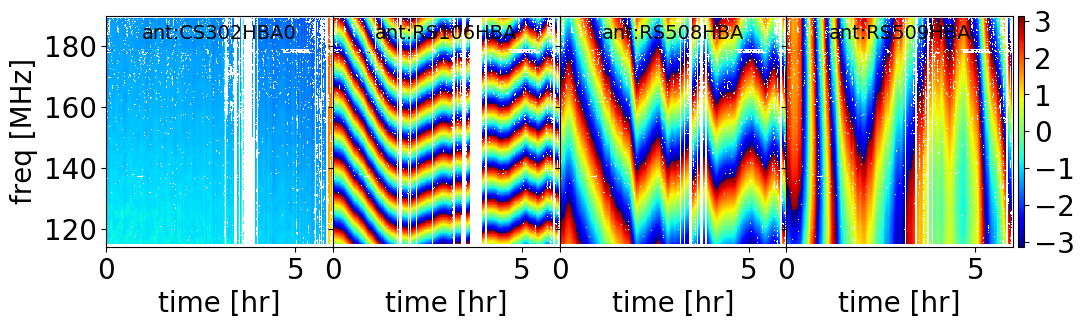}\\
 \includegraphics[width=\textwidth]{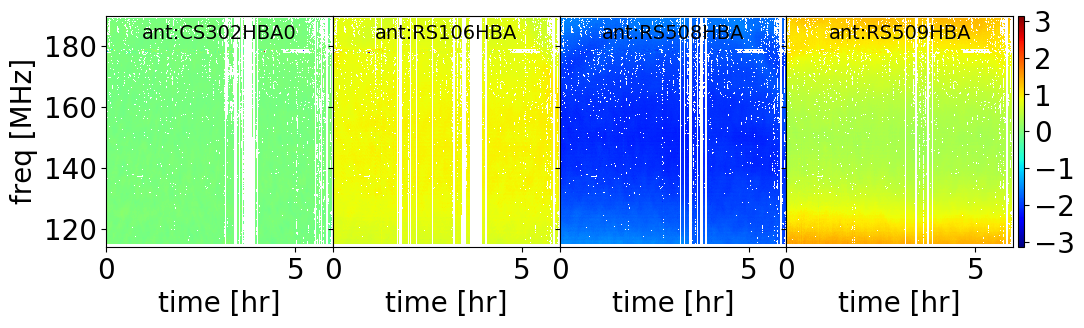}
 \caption{Phase solutions, same as in Fig.~\ref{fig:ph_LBA} but for HBA. White regions are flagged data.}
 \label{fig:ph_HBA}
\end{figure*}

\begin{figure*}[ht!]
\centering
 \includegraphics[width=\textwidth]{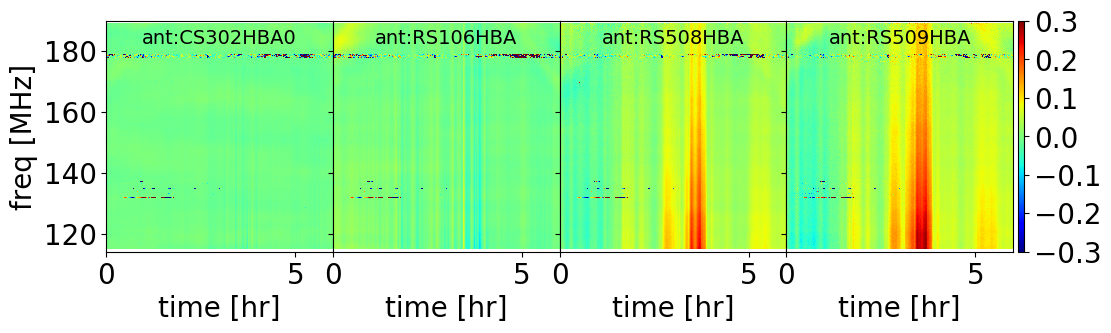}
 \caption{Solution for the rotation matrix, same as in Fig.~\ref{fig:rot_LBA} but for HBA.}
 \label{fig:rot_HBA}
\end{figure*}

\begin{figure*}[ht!]
\centering
 \includegraphics[width=\textwidth]{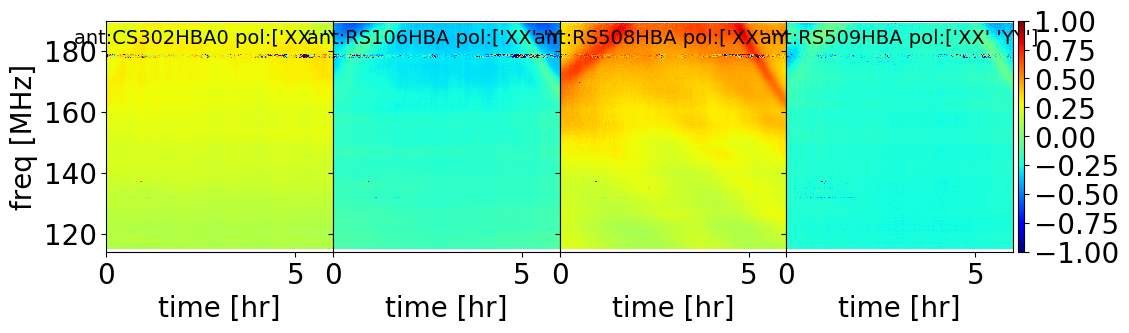}\\
 \includegraphics[width=\textwidth]{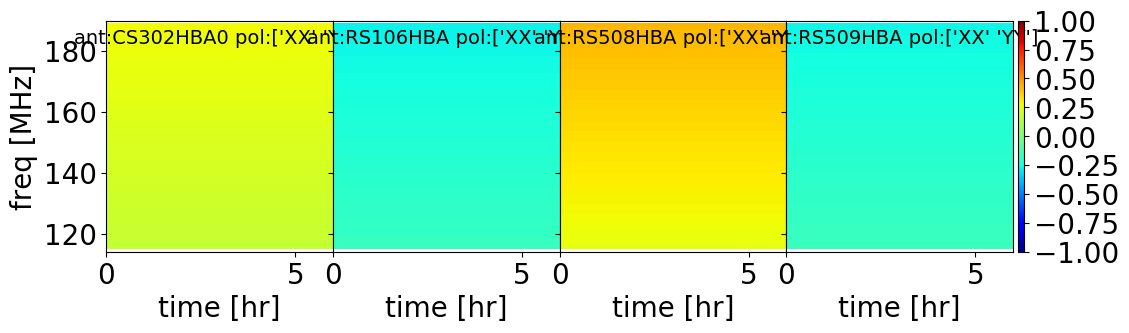}\\
 \includegraphics[width=\textwidth]{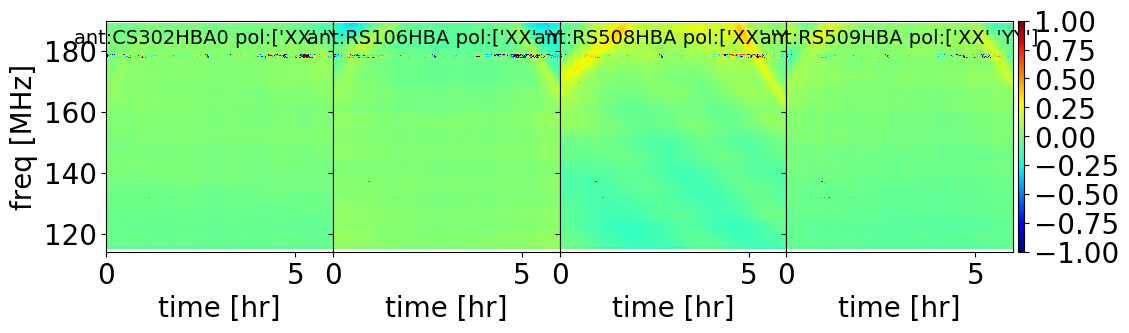}
 \caption{Differential phase solutions, same as in Fig.~\ref{fig:pa_LBA} but for HBA.}
 \label{fig:pa_HBA}
\end{figure*}

\begin{figure*}[ht!]
\centering
 \includegraphics[width=\textwidth]{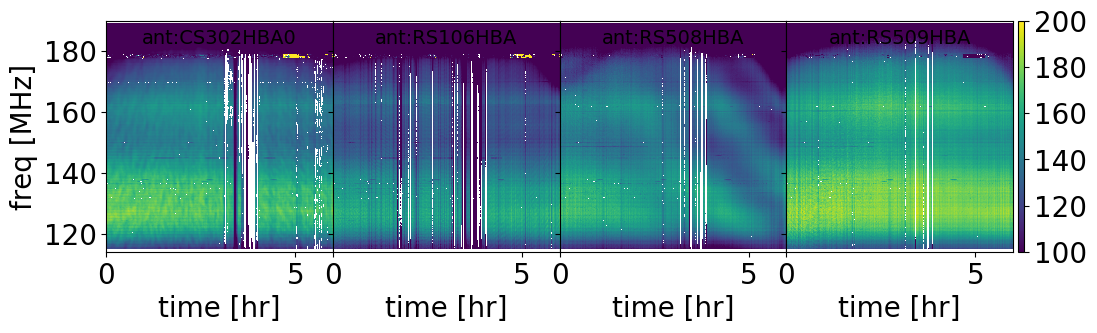}\\
 \includegraphics[width=\textwidth]{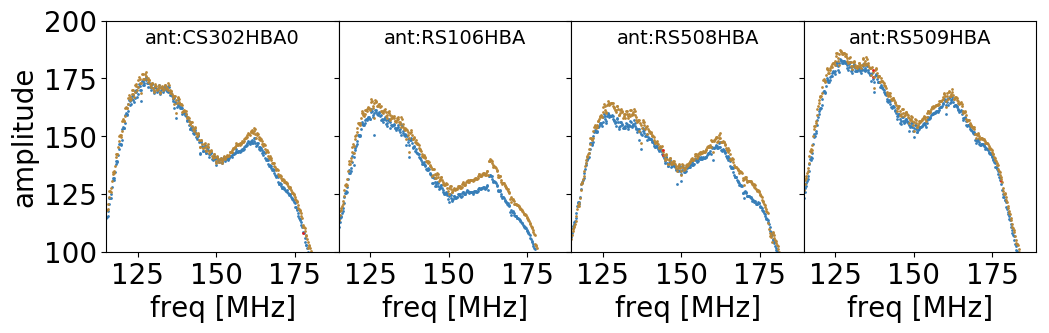}\\
 \includegraphics[width=\textwidth]{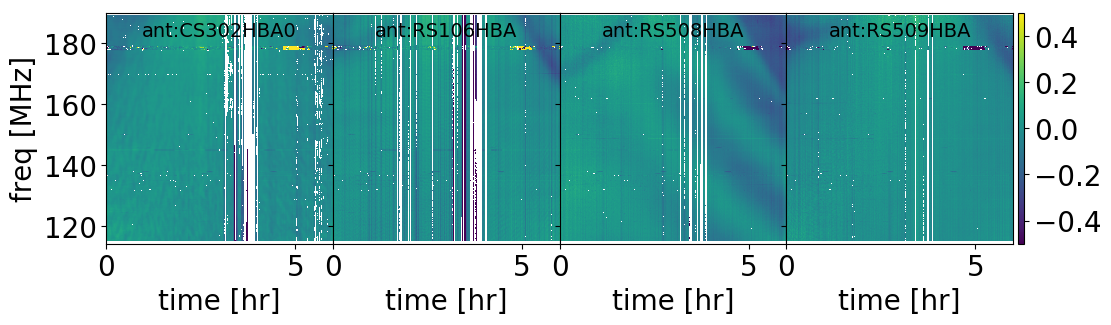}
 \caption{Amplitude solutions, same as in Fig.~\ref{fig:amp_LBA} but for HBA.}
 \label{fig:amp_HBA}
\end{figure*}

\begin{figure*}[ht!]
\centering
 \includegraphics[width=0.32\textwidth]{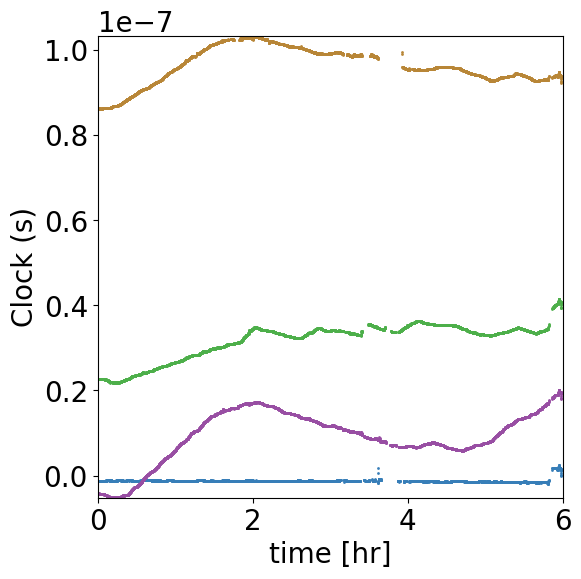}
 \includegraphics[width=0.33\textwidth]{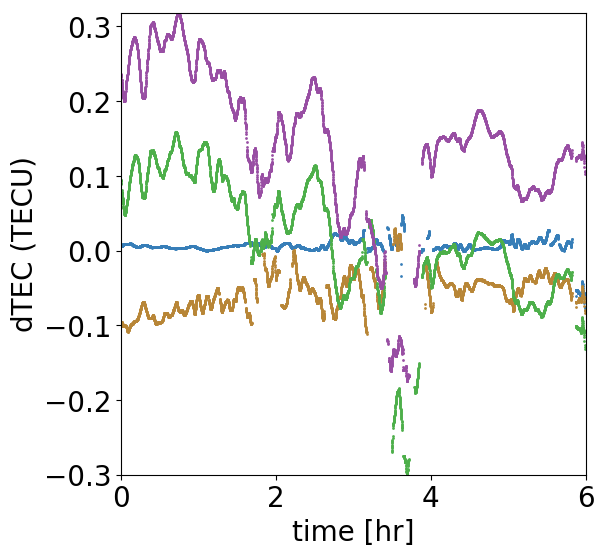}
 \includegraphics[width=0.33\textwidth]{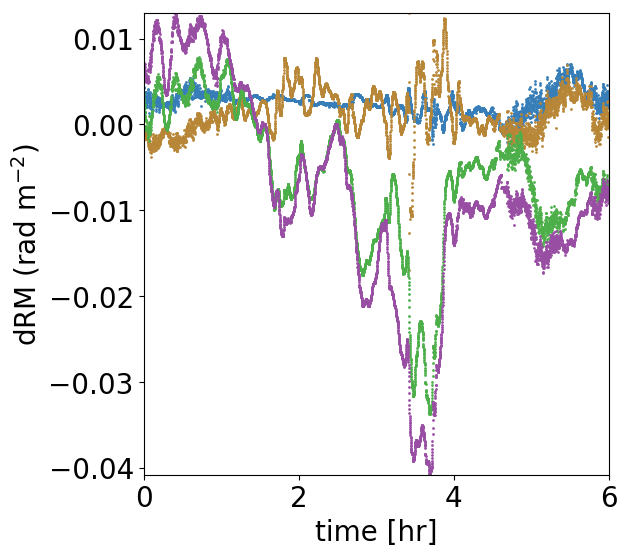}
 \caption{Same as in Fig.~\ref{fig:syseff_LBA} but for HBA. Owing to substantial flagging in the data set the clock/TEC separation procedure produced a few jumps in the output data streams.}
 \label{fig:syseff_HBA}
\end{figure*}

We report the plots of the solutions for the HBA data set. Each figure corresponds to one of LBA solutions presented in Sec.~\ref{sec:strategy}.

\newpage

\section{Calibration in DPPP}
\label{sec:dppp}

The low-level calibration routines are implemented in DPPP \citep{VanDiepen2018}. This software package is written to perform operations on visibilities measurement sets \citep{VanDiepen2015} while iterating over data in time order. The DPPP tool reads every chunk of data once, then processes a configurable list of operations (steps) on each chunk, before writing it back to disc. In this way, disc input/output is minimised. This is particularly useful for I/O limited operations.

We implemented a DPPP step, \texttt{gaincal}, to perform calibration following the algorithm from \cite{Mitchell2008, Salvini2014a}. We extended the algorithm to find one solution for many channels or many time slots, by treating all visibilities from a channel / time slot as a new sample of the coherencies. Our implementation supports scalar solutions, diagonal solutions (separate solutions for X- and Y-dipoles), and full-Jones solutions \citep{Salvini2014b}. The full set of options is described in the on-line documentation\footnote{\url{https://www.astron.nl/lofarwiki/doku.php?id=public:user_software:documentation:ndppp}}.

To optimise the computation, we temporarily stored the visibilities in a full matrix. A particular order of the various axes was chosen to make memory access linear for all use cases. This enables the compiler to vectorise the code. The order in which the correlations between stations (each with polarisations X and Y) are stored is, from slow to fast varying as follows : station 1, polarisation 1, channel, time, polarisation 2, and station 2.

Constrained solutions are possible by inserting a constraining step between the iterations of gaincal. For example, dividing out the amplitude at every iteration yields an optimal phase-only solution (as was already mentioned in \cite{Salvini2014a}). We implemented several new constraints using a new constraint framework (Offringa et al, in preparation).

We implemented a new constraint for obtaining solutions of the form
\begin{align}
\operatorname{diag}(g_{00}, g_{11}) \cdot \operatorname{Rot}(\phi) = \big(\begin{smallmatrix} g_{00} & 0 \\ 0 & g_{11}\end{smallmatrix}\big) \big(\begin{smallmatrix}\cos(\phi) & -\sin(\phi) \\ \sin(\phi) & \cos(\phi)\end{smallmatrix}\big), \quad\text{for $\phi\in[-\pi/2, \pi/2),$}
\end{align}
where $g_{00}$, $g_{11} \in \mathbb{C}$ are the gain solutions for the X and Y dipole, respectively. This represents calibration for a rotation of the orthogonal dipoles and a separate gain for the X and Y dipoles. This constraint works by iterating in the full-Jones gaincal algorithm, and in each step constraining the iterand $G=\big(\begin{smallmatrix}g_{00} & g_{01}\\g_{10}&g_{11}\end{smallmatrix}\big)$ to the mentioned form.

Constraining the iterand is done by first finding the best-fit rotation $\phi^0$. For a given full Jones solution iterand $G$, this is given by
\begin{align}
\begin{split}
\phi^0\Big[\big(\begin{smallmatrix}g_{00} & g_{01}\\g_{10}&g_{11}\end{smallmatrix}\big)\Big] = &\tfrac{1}{2}\arg(g_{00}+g_{11}-g_{01}i+g_{10}i) - \\
& \tfrac{1}{2}\arg(g_{00}+g_{11}+g_{01}i-g_{10}i) + k\pi,
\end{split}
\end{align}
where $k\in\mathbb{Z}$ is chosen such that $\phi^0\in[-\pi/2,\pi/2)$.

Verifying this, indeed we can extract a given rotation $\phi$, independent of the diagonal terms as follows:
\begin{align*}
\phi^0(\operatorname{diag}(g_{00},g_{11})) \cdot \operatorname{Rot}(\phi)=\\
\tfrac{1}{2} \arg\big((g_{00}+g_{11})(\cos\phi+i\sin\phi)\big)-\tfrac{1}{2}\arg\big((g_{00}+g_{11})(\cos\phi-i\sin\phi)\big) = \\
\tfrac{1}{2} \Big[\arg(g_{00}+g_{11}) + \arg(e^{i\phi})\Big] -
\tfrac{1}{2} \Big[\arg(g_{00}+g_{11}) + \arg(e^{-i\phi})\Big] = \\
\tfrac{1}{2} \big(\arg(e^{i\phi}) - \arg(e^{-i\phi}\big) = \phi
\end{align*}
The terms $g_{00}^0$, $g_{11}^0$ are found from $\big(\begin{smallmatrix}g_{00}^0 & 0 \\ 0 & g_{11}^0\end{smallmatrix}\big) =\operatorname{diag}(G) \cdot \operatorname{Rot}(-\phi^0)$.

In the presence of white noise, since all used operations are linear, this extracts the best-fit rotation and diagonal terms.

\section{LOFAR Solution Tool (LoSoTo)}
\label{sec:losoto}

The LOFAR Solution Tool (LoSoTo) is a Python package that handles radio calibration solutions in a variety of ways. The data files used by LoSoTo are called H5parm and are based on the HDF5 standard\footnote{\url{http://www.hdfgroup.org/HDF5/}}. Current LOFAR software is able to read/write solutions in such data file format.

\subsection{H5parm format}
H5parm is simply a list of rules that specify how data are stored in an HDF5 file. The H5parm format relates to HDF5 in the same way that CASA solutions tables relates to MeasurementSet \citep{VanDiepen2015}. As an open source project developed by a large community of people, the HDF5  has some very easy-to-use Python interfaces (e.g. the \texttt{pytables} module). The LoSoTo package stores solutions in arrays organised in a hierarchical fashion. This provides enough flexibility but preserves performance. Solutions of multiple data sets can be stored in the same H5parm (e.g. the calibrator and target field solutions of the same observation) into different solution sets (solset). Each solset can be seen as a container for a logically related group of solution tables (soltab). Each solset contains an arbitrary number of soltabs plus some tables with metadata on antenna locations and pointing directions. Soltabs can have an arbitrary name and they are in turn containers: inside each soltab there are several arrays that are the real data holders. Typically, there are a number of one-dimensional arrays storing the axes values and two $n$-dimensional (where $n$ is the number of axes) arrays, ``values'' and ``weights'', which contain the solution values and the relative weights. By convention, a weight of zero means a flagged solution. Soltabs can have an arbitrary number of axes of arbitrary data type. We list some examples of common soltabs and possible axes:

\begin{itemize}
 \item amplitudes: time, freq, pol, dir, ant
 \item phases: time, freq, pol, dir, ant
 \item clock: time, ant, [pol]
 \item tec: time, ant, dir, [pol]
\end{itemize}

Theoretically the value and weight arrays can only be partially populated, leaving NaNs (with 0 weight) in the gaps. The main benefit of this is that it enables different time resolutions for different antennas, at the cost of an increment of the data size. H5parm can be compressed using a number of algorithms, this reduces the data size but increases the reading and writing time.

\subsection{LoSoTo}

The LoSoTo packaged can be used to perform a series of operations on a specified H5parm. The code receives its commands by reading a parset file. Alternatively, any operation can be called using a python interface. Subsets of data can be selected for each operation using lists of axes values, regular expressions, or intervals. These are the operations that LoSoTo can currently perform:

\begin{description}
 \item[ABS] Take absolute value.
 \item[CLIP] Clip solutions around the median.
 \item[CLOCKTEC] Separate phase solutions into clock and TEC (1st and 3rd order). The clock and TEC values are stored in output soltabs with type: clock, tec, and tec3rd.
 \item[DIRECTIONSCREEN] Fit spatial screens to solutions of multiple stations.
 \item[DUPLICATE] Duplicate a table.
 \item[FARADAY] Faraday rotation extraction from RR/LL phase solutions or a rotation matrix.
 \item[FLAGEXTEND] For each datum check if the surrounding data are flagged to a certain percentage (in multi-dimensional space), then decide whether to flag that datum as well.
 \item[FLAG] An outlier flagging procedure.
 \item[NORM] Normalise the solutions to a given value.
 \item[PLOT] Advanced plotting routine (solution plots in this paper were created with this operation).
 \item[POLALIGN] Estimate polarisation misalignment as a delay.
 \item[RESET] Reset all the selected amplitudes to 1 and all other selected solution types to 0.
 \item[RESIDUALS] Subtract/divide two tables or remove a clock/tec/tec3rd/rotation measure effect from a phase table.
 \item[REWEIGHT] Modify the weights by hand.
 \item[SMOOTH] A smoothing function: running median on an arbitrary number of axes, running polyfit on one axis, or set all solutions to mean/median value.
 \item[STATIONSCREEN] Fit spatial screens to solutions of a single station.
 \item[STRUCTURE] Calculate the ionospheric structure function.
 \item[TEC] Estimate TEC using a brute force fit on phase solutions.
\end{description}

The code is still under development and new operations are expected to be added in the future. The code is freely available at \url{https://github.com/revoltek/losoto}. Documentation and examples are also present at that website.
\end{appendix}

\end{document}